\documentclass{WileyMSP-template}
\usepackage[usenames,dvipsnames]{xcolor}
\usepackage{amsmath,amsfonts}
\usepackage{bm}
\definecolor{oceanboatblue}{rgb}{0.0, 0.47, 0.75}
\definecolor{orange}{rgb}{1,0.5,0}
\definecolor{goodgreen}{rgb}{0.1,0.5,0}
\definecolor{goodred}{rgb}{0.7,0,0}
\usepackage[colorlinks,urlcolor=goodgreen,citecolor=blue,linkcolor=goodred]{hyperref}
\usepackage{subcaption}

\usepackage{lineno}

\begin{document}

\pagestyle{fancy}
\rhead{\includegraphics[width=2.5cm]{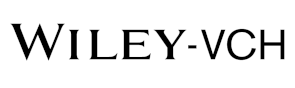}}

\title{\textbf{Superconducting proximity effect in $d$-wave cuprate/ graphene heterostructures}}

\maketitle

\author{David Perconte}
\author{Dario Bercioux}
\author{Bruno Dlubak }
\author{Pierre Seneor}
\author{F. Sebastian Bergeret}
\author{Javier E. Villegas*}

\begin{affiliations}
David Perconte\\
Institut Néel,\\
Univ. Grenoble Alpes,\\ 
CNRS, Grenoble INP,\\
25 avenue des martyrs,\\
38000 Grenoble, France\\

Dario Bercioux\\
Donostia International Physics Center, 
20018 Donostia-San Sebasti\'an, \\
Spain \& \\
IKERBASQUE, Basque Foundation for Science, \\
Euskadi Plaza, 5, \\
48009 Bilbao, Spain\\

Bruno Dlubak, \\
Pierre Seneor,\\
Unit\'e Mixte de Physique,
CNRS, Thales, 
Universit\'e Paris Saclay,\\
91767, Palaiseau, France\\

F. Sebastian Bergeret\\
Centro de F\'isica de Materiales (CFM-MPC) Centro Mixto CSIC-UPV/EHU,\\
20018 Donostia-San Sebasti\'an, \\
Basque Country, Spain \& \\ 
Donostia International Physics Center, 
20018 Donostia-San Sebasti\'an, \\
Spain\\

Javier E. Villegas,\\
Unit\'e Mixte de Physique,
CNRS, Thales, 
Universit\'e Paris Saclay,\\
91767, Palaiseau, France\\
javier.villegas@cnrs-thales.fr
\end{affiliations}


\justifying
\begin{abstract}

Superconducting proximity effects in graphene have received a great deal of attention for over a decade now. This has unveiled a plethora of exotic effects linked to the specificities of graphene’s electronic properties. The vast majority of the related studies are based on conventional, low-temperature superconducting metals with isotropic $s$-wave pairing. Here we review recent advances made on the less studied case of unconventional high-temperature superconducting cuprates. These are characterized by an anisotropic $d$-wave pairing, whose interplay with Dirac electrons yields very rich physics and novel proximity behaviours. We provide a theoretical analysis and summarize the experiments reported so far. These unveil hints of proximity-induced unconventional pairing and demonstrate the gate-tunable, long-range propagation of high-temperature superconducting correlations in graphene. Finally, the fundamental and technological opportunities brought by the theoretical and experimental advances are discussed, together with the interest in extending similar studies to other Dirac materials. 
\end{abstract}

\section{Introduction}
Proximity effects in graphene have been intensively investigated ever since the early studies evidenced that its electronic properties~\cite{Castro_Neto_2009} transform both the underlying physical mechanisms~\cite{Beenakker_2008} and the related phenomenology~\cite{Heersche_2007}. As we discuss in Section~\ref{2D_material},  most of the specificities are linked to the fact that graphene is a two-dimensional (2D) Dirac material, which in many situations can be described as a gapless semiconductors in which carriers exhibit a linear dispersion around zero energy. In addition, because it presents a relatively low carrier concentration, electrostatic gating allows large modulations of the electrochemical potential. This has a strong impact on the mechanism underlying the propagation of superconducting correlations into graphene, the Andreev reflection, and thus provides a knob for tuning the proximity effect. Beyond this facet, other characteristics such as the possibility to induce a transition from bulk to edge transport~\cite{Allen_2015,Ben_Shalom_2015,Amet_2016} have fostered significant research efforts.  The vast amount of work has been based on conventional $s$-wave superconductors and, because it is out of the scope of this review, it will be only briefly recalled hereafter.
On the other hand, superconducting cuprates present many properties that make then attractive in the framework of superconducting proximity in graphene. One can start by mentioning the most obvious one, which is that their high critical temperature is appealing for technological applications. For example, gate-tunable Josephson coupling across graphene, which has been demonstrated with low-$T_\text{C}$ superconductors, would be a game-changing asset in the growing field of high-temperature superconducting electronics~\cite{Ortlepp2006,Koelle1999a,Ouanani2016,Swiecicki2012}. Beyond that, arguably one of the most appealing properties of the cuprates is that they present anisotropic $d$-wave pairing. This means that the amplitude of the superconducting gap and the superconducting phase vary in the momentum space, particularly in the ($k_x$-$k_y$)-plane of graphene’s 2D Dirac electrons. As discussed in Section~\ref{Theory_dwave_graphene}, this has a strong influence on the proximity effect, for example on the Andreev reflection process and on the conductance across the cuprate/graphene interface.
Along with their interest and the unique opportunities they offer, cuprate superconductors bring complexity, particularly regarding the fabrication of solid-state devices, which is much more challenging than with conventional superconductors. However, as we detail below, \emph{ad hoc} techniques have recently allowed relieving some of those constraints, enabling the experiential demonstration of the proximity effect in graphene in contact with superconducting cuprates~\cite{Di_Bernardo_2017,Perconte_2017,Perconte2020} as well as proximity between graphite and BSCCO~\cite{Zareapour2017}. 

The above ideas are unfolded hereafter, the review being organized as follows: the specificities of graphene and $d$-wave superconductors are summarized in Section~\ref{2D_material} and~\ref{Dwave_material}, where in addition the motivation for mixing both types of materials in proximity devices is rationalized. A theoretical section follows (Section~\ref{Theory}), which first focuses on the particularities of the proximity effect between $d$-wave superconductors and trivial metals (Section~\ref{Theory_dwave_metal}) and then on those of the proximity between Dirac materials and conventional superconductors (Section~\ref{sec_s_graphene}). These two theoretical sections are introductory to that devoted to the main topic of the review, which is $d$-wave/Dirac materials interfaces (Section~\ref{Theory_dwave_graphene}), with a stronger focus on the case of graphene and an extension to the surface state of 3D topological insulators.  Section~\ref{Experiment} is devoted to the experiments. After briefly summarizing some of the work based on $s$-wave superconductors (Section~\ref{exp_swave}), we focus on the scanning tunneling microscopy studies and on devices based on $d$-wave/graphene heterostructures (Sections~\ref{STM} and~\ref{Device_Fabrication}). A brief discussion of experiments involving $d$-wave cuprates and other Dirac materials is presented thereafter (Section~\ref{exp_TI}), which is followed by the concluding remarks and outlook (Section~\ref{Conclusion}).

\subsection{Two-dimensional Dirac materials: graphene and beyond}\label{2D_material}

\begin{figure}
    \centering
    \includegraphics[width=0.95\textwidth]{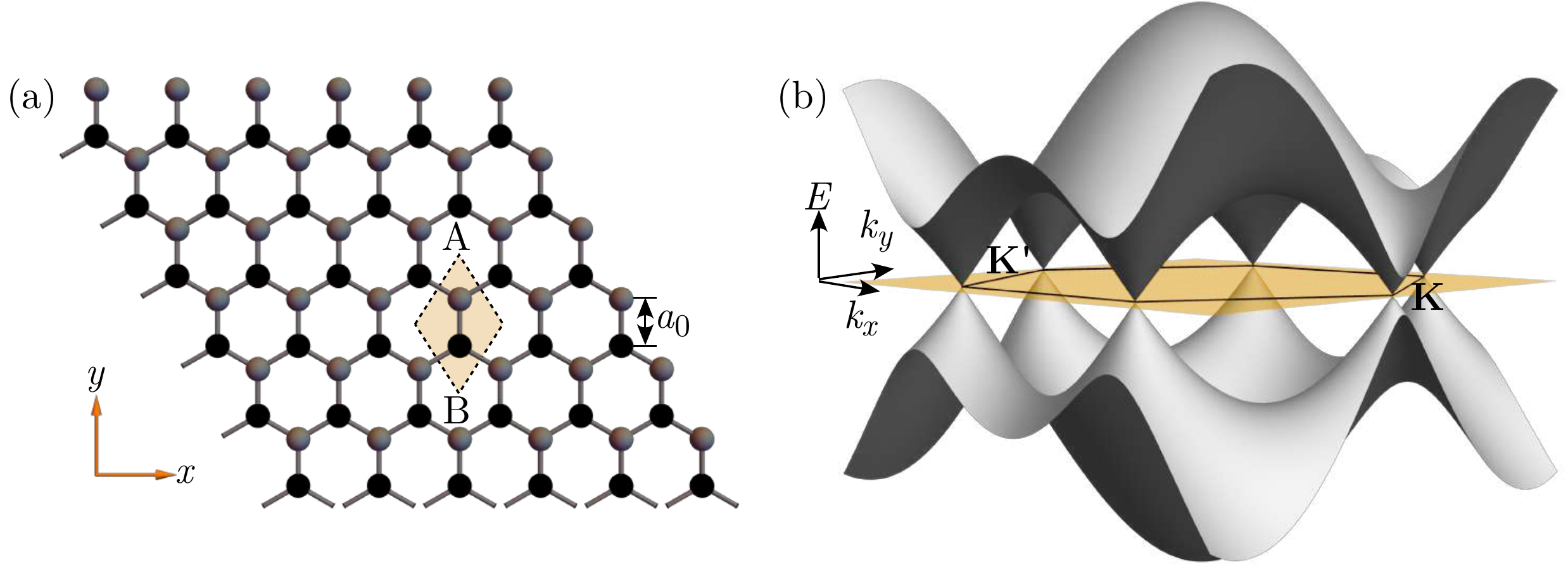}
    \caption{\label{fig_one} (a) Real-space structure of the graphene lattice, the gray and the black spheres represent the two inequivalent carbon atoms in the unit cell of graphene denoted by A and B (yellow region). The carbon-carbon distance is indicated via $a_0$. (b) The band structure of the graphene.  The hexagon in the yellow plane represents the first Brillouin zone. The  ``Dirac'' points are given by the touching points of the valence and conduction bands in the six corner of the hexagon, by symmetry only two of those belong to the first Brillouin zone and are usually named $\mathbf{K}$ and $\mathbf{K'}$.}
\end{figure}

Graphene, an atom thin hexagonal crystal of carbon, is the most archetypal two-dimensional (2D) material~\cite{Novoselov2005}, see Fig.~\ref{fig_one}(a). It presents remarkable structural, optical and electronic properties~\cite{Ferrari2015}. Thanks to its unique characteristics, graphene has become a superlative material: when compared to its size it is probably the strongest material per unit size (tensile strength in the 100~Gpa range~\cite{Gong2012}, on par with diamond and far greater than steel) while keeping striking elastic properties (Young’s Modulus in the TPa range~\cite{Lee2008}), it offers the highest thermal conductivity at room temperature (in the order of $5000$~W/mK~\cite{Balandin2008}), it has a large surface area of $2630$~m$^2$/g~\cite{Zhu2010}, absorbs only 2.3 \% of light~\cite{Nair2008} and is impervious so that even helium can't pass through~\cite{Bunch2008}. But it is its remarkable electronic and transport properties that have initially drawn the most attention~\cite{Castro_Neto_2009,Goerbig_2011}. As a matter of fact, the Dirac band structure is one of the striking features of graphene, it was early unveiled by Wallace sixty years ago~\cite{Wallace_1947} to present a linear dispersion crossing at the so-called Dirac points --- see Fig.~\ref{fig_one}(b). This has led to impressive mobilities, robustness of its 2D electron gas and exotic transport phenomena such as massless quasiparticles, chirality, uncommon Landau levels splitting and the Klein tunneling paradox~\cite{Torres_2013}.
%
%
%
%
In terms of electronic structure, the carbon atoms in the honeycomb structure of graphene form the $\sigma$-bonds with neighboring carbon atoms through $sp_2$ orbitals while the remaining $p_z$ ones overlap to form a band of filled $\pi$ orbitals and a band of empty $\pi^*$ orbitals that give rise to respectively the valence band and the conduction band. Thanks to this electronic configuration, graphene can be thought as a two-dimensional zero-gap semimetal with in-plane conductivity --- see Fig.~\ref{fig_one}(b). The electronic band structure of graphene directly arises from the two inequivalent carbon atoms of the graphene crystal structure, see Fig.~\ref{fig_one}(a). This also gives rise to a key feature of graphene for electronic transport: its low-energy quasiparticles formally described by the Dirac-like Hamiltonian $\mathcal{H}_\text{G}=v_\text{F} \bm{\sigma}\cdot\bm{p}$, where $v_\text{F} = 10^6 m/s$ is the Fermi velocity, $\bm{\sigma}\equiv(\sigma_x,\sigma_y)$ are the Pauli matrices corresponding to a so-called pseudo-spin associated to the two carbon atoms in the lattice unit cell, and $\bm{p}\equiv(p_x,p_y)$ is vector of the momenta operators in the graphene plane. The two energy bands ($\pi,\pi^*$) intersect at the corner of the hexagonal first Brillouin zone yielding the conical energy spectrum at the $\mathbf{K}$ and $\mathbf{K}'$ ``Dirac" points. Hence, quasiparticles in graphene exhibit the linear dispersion relation $E = \hbar v_\text{F} k$ as if they were massless relativistic particles, only with a Fermi velocity $v_\text{F} $ approximately 300 times slower than the speed of light. The fact that charge carriers in graphene are described by the Dirac-like equation (\emph{i.e.} linear dispersion relation) rather than the usual Schr\"odinger equation (\emph{i.e.} free electron with parabolic-like dispersion relation)  has direct consequence on the transport properties when contacted with superconducting electrodes, as discussed  in Section~\ref{sec_s_graphene}. The difference is particularly striking when considering effects with energies close to the crossing point of the linear dispersion -the charge neutrality point. Only in this regime, a clear distinction between linear and parabolic energy dispersion plays a significant role. 

Still, while theoretically unveiled more than sixty years ago, experimentally, graphene has been first isolated in 2004~\cite{Novoselov2004}.  By leveraging on the exfoliation technique, it has been possible to access relatively easily high-quality crystals and explore their unique properties. As early as in 2004, the mobility of graphene was measured to be exceptionally high. For mechanically exfoliated graphene, mobilities $\mu>10000$~cm$^2$/Vs at room temperature are now currently reported, for carrier concentrations (electrons or holes) that can be modulated by a standard gate up to high values of $10^{13}$cm$^{-2}$~\cite{Chen2008,Morozov2008,Du2008,Novoselov2004}. The intrinsic (low-temperature) limit of graphene mobility was revealed by Landau spectroscopy measurements on high purity monolayers on the surface of a graphite crystal: the reported mobility values exceeded $10^7$~cm$^2$/Vs~\cite{Neugebauer2009}.

The main difficulty linked to mechanically exfoliated graphene has however concerned the small size of the obtained sheets and their random distribution on the substrate surface~\cite{Mohiuddin2008}. An additional but fundamental constrain has been the necessity of relying on a specifically tailored substrate (usually Si/SiO$_2$) leading to interferences of Fabry-P\'erot type and allowing their observation through contrast enhancement~\cite{Blake2007}. Those limitations, which would be prohibitive for any exploitation that requires a large-scale integration of components, have also made any systematic study and integration with specific substrates more difficult. As such, the implementation of graphene has been a challenge and the most important problem has been lying in the preparation of high-quality and well-defined graphene over large surfaces. Thus, many material studies have aimed at producing large area graphene (\emph{i.e.} substrate size surfaces)~\cite{Berger2004,Obraztsov2009}. In this direction, chemical vapor deposition (CVD) has quickly appeared as a very relevant solution towards the production of monolayers on wafer scales~\cite{Chen2011,Bae2010,Kidambi_2012,Weatherup2012}. Layers of high-quality over large surfaces are now achieved and easily transferred to any substrate of choice in soft ambient conditions~\cite{Mzali_2016,Kang2012}. This has allowed to envision the integration of graphene in many systems, even the most delicate or specific ones and has been key to the integration of graphene with advanced complex oxides such as high-$T_\text{C}$ superconductors~\cite{Perconte_2017}.

The broader Dirac-material family of three-dimensional topological insulators (3DTI) shares part of the electronic properties of graphene. These materials are characterized by a strong spin-orbit coupling, they present an insulating bulk, but have conducting surface states~\cite{Wehling_2014}. As in standard insulators, in bulk, a finite size energy gap separates their valence and conduction bands. However, their surface or any boundary with a normal insulator or the vacuum is characterized by surface states that can be described by an effective Dirac-like Hamiltonian. These surface states live inside the bulk energy gap and are topologically protected unless time-reversal symmetry is broken. For example, strained HgTe represents an important platform for 3DTI. Additionally, 3DTIs are hosted in several layered structures based on Sb, Bi, Se and Te, as Bi$_2$Se$_3$, Bi$_2$Te$_3$ and Sb$_2$Te$_3$. These are multi-layers systems that crystallize in tetradymite structure. Thus, they consist of covalently bonded quintuple layers (\emph{e.g.}, Se?Bi?Se?Bi?Se) that are stacked in ?A?B?C?A?B?C? manner and are weakly interacting with van der Waals force~\cite{Ando_2013}.

The effective Hamiltonian describing these systems is similar to the graphene one but without the valley degree-of-freedom. Additionally, the pseudo-spin $\bm{\sigma}$ is substituted by the fermionic spin degree-of-freedom $\bm{s}\equiv(s_x,s_y)$, thus their effective low-energy Hamiltonian is $\mathcal{H}_\text{3DTI}=v_\text{F} (\bm{s}\times\bm{\nabla})_z$.
Additional cubic terms in momentum to this effective Hamiltonian can describe the high-energy hexagonal warping of the Dirac cone of these layered 3DTI systems~\cite{fu_2009}. This effective Hamiltonian encodes a peculiar characteristic of 3DTI that is the \emph{spin-momentum locking}, namely the spin degeneracy is lifted off for the surface states and the direction of the spin is always perpendicular to the motion direction~\cite{Bardarson_2013}, this property being similar to the photon helicity.  The spin-momentum locking is typical of all the systems with a strong spin-orbit interaction~\cite{Bercioux_2015}. The effective Hamiltonian $\mathcal{H}_\text{3DTI}$ describes the electron gas on a single 3DTI surface, a similar Hamiltonian describes the opposite surface with carriers with opposite helicity. However, in transport experiments only one of these surfaces is contacted, thus a model accounting for a single Dirac cone is sufficient for our purposes.

Since 2010, many new families of 3DTI have been uncovered and studied, expanding the available exotic transport properties of graphene~\cite{Ando_2013,Vergniory_2019}. The advantage of the binary compounds as Bi$_2$Se$_3$ is that they can be easily prepared under several forms: bulk single crystals, thin films, nanobelts, or nanoplates. However, they all presents a high density of defects. 
From the experimental point of view, one of the major challenges for designing quantum transport experiments is related to address only the charge carriers on a specific surface. The problem arises from a small but finite coupling between the surface and the bulk states of the 3DTI materials. This problem is usually addressed by introducing more complex \emph{ad hoc} two-band models, one describing the surface state, similar to $\mathcal{H}_\text{3DTI}$, and one accounting for the electronic bulk states~\cite{Ando_2013}. A way out from avoiding the influence of the bulk states on the topological surface modes is to move from binary compounds (\emph{e.g.} Bi$_2$Se$_3$) to ternary compounds~\cite{Ando_2013,Yang_2013,Culcer_2020}. For example, we can consider (Bi$_{1-x}$Sb$_x$)$_2$Te$_3$ where the chemical potential can has been tuned from bulk to valence conduction changing the Sb concentration~\cite{Zhang_2011,Kong_2011}. 
There is another property of interest of 3DTIs is connected to the spin-momentum locking: the surface states of 3DTIs  present a particular photo-response to circularly-polarized light. However, the associated spin-polarized current is not detected so easily. The problem arises because spin-momentum locking makes the charge and the spin life-times identical, as a consequence, the mean-free-path and the spin diffusion length are of the same magnitude. Therefore, in diffusive transport regimes, the spin polarization is decreased. The only way to prevent this problem is to perform experiments in the ballistic regime~\cite{Burkov_2010,Mellnik_2014}.

\subsection{$d$-wave cuprates: unique properties and opportunities for proximity effects in Dirac materials}\label{Dwave_material}

Cuprates constitute a large family of superconducting oxides that have remained a central problem in condensed-matter physics since their discovery over three decades ago. The origin of superconductivity, which in these materials is far from being understood~\cite{Zhou2021}, is connected to  the presence of competing electronic ground-states~\cite{Proust2019} that stem from strong correlations and yield a very rich phase diagram~\cite{Keimer2015a}. These matters are the focus of intense research efforts that are out of this review’s scope.
%
%
\begin{figure}
    \centering
    \includegraphics[width=0.95\textwidth]{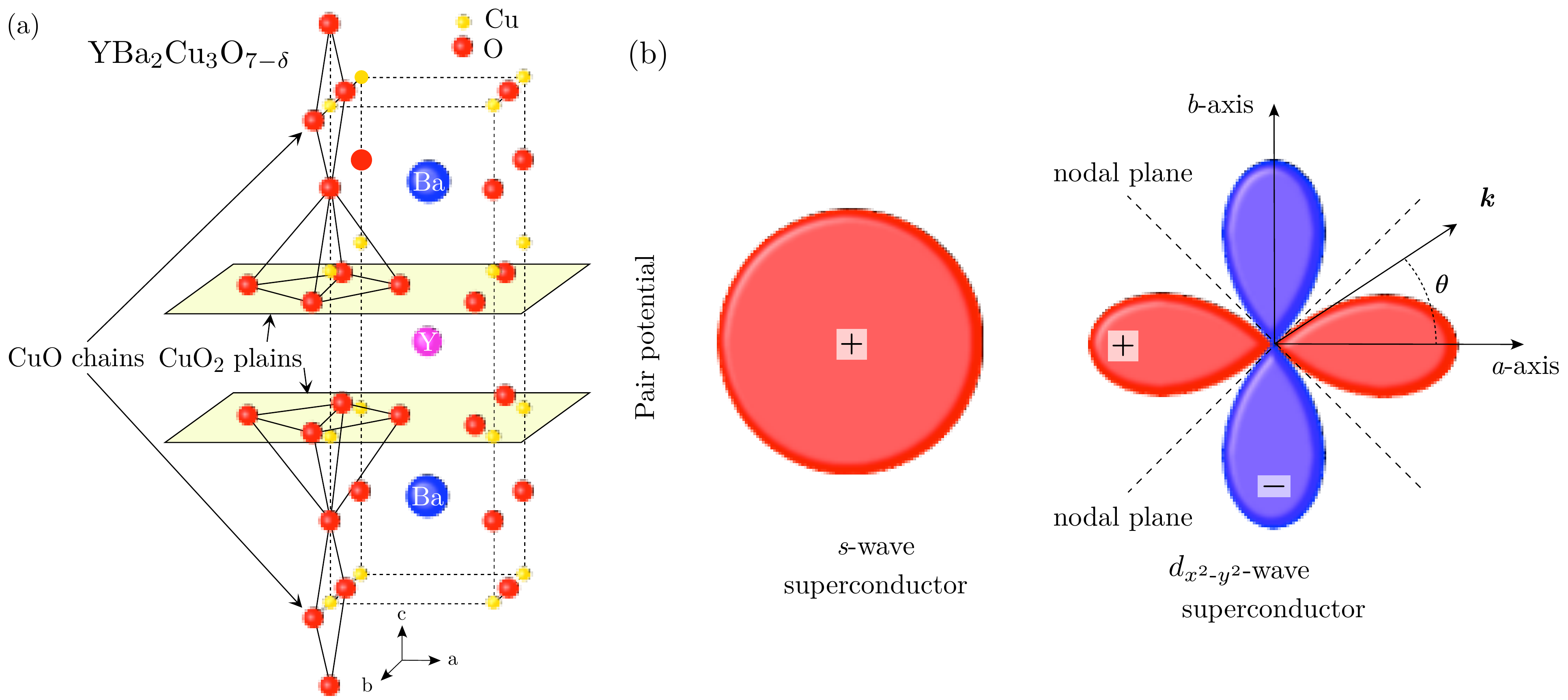}
    \caption{\label{fig:YBCO}Archetypal $d$-wave superconductor. (a) Unit cell of the high-temperature superconducting cuprate YBa$_2$Cu$_3$O$_{7-\delta}$~\cite{Hott2016}. (b) Comparison between conventional $s$-wave pairing and the $d$-wave pairing characteristic of most cuprate superconductors. In the latter, the order parameter is anisotropic, its amplitude being dependent on the direction as shown in the right picture, and the superconducting phase is shifted by $\pi$ between $d$-wave lobes~\cite{Kashiwaya2000}.}
\end{figure}
%
%
Along with those electronic specificities, cuprates also present very unique superconducting properties. For many of the compounds the superconducting critical temperature $T_\text{C}$ exceeds the normal boiling point of nitrogen which, accompanied by large critical currents and fields, confers to the cuprates much potential for a variety technological applications~\cite{Rogalla2011}. These span from high-current coated conductors~\cite{MacManus-Driscoll2021}, which can be used for example for the generation of high magnetic fields~\cite{Gupta_2018}, to high sensitivity sensors such as bolometers~\cite{Lee1996,Kokkoniemi2020} and transition edge detectors~\cite{Ullom2015a} or quantum interference devices~\cite{Koelle1999a,Crete2020} such as quantum antennas~\cite{Ouanani2016} and magnetometric sensors~\cite{Couedo2019}.
Another key property most cuprates share is that they exhibit a $d$-wave pairing. This means  that their superconducting gap is anisotropic, with respect to  the  direction  of  the momentum and  also has a change of sign, see Fig.~\ref{fig:YBCO}(b),  Refs.~\cite{Shen1993,Aubin1997,Kirtley2006}. This results in a high-density of zero-energy quasiparticle (Andreev) surface bound-states whose density depends on the orientation of the superconductor's surface relative to the crystallographic axes~\cite{Kashiwaya_1995,Kashiwaya2000}.
The surface Andreev bound states result in a zero-energy peak in the tunneling conductance that sharpens upon decreasing the temperature~\cite{Kashiwaya_1995,Kashiwaya2000}. As it will be discussed in Section~\ref{Theory}, this has strong effects on the conductance across $d$-wave/metal interfaces in general, and across $d$-wave/graphene interfaces in particular. The characteristic $\pi$ phase shift between $d$-wave lobes~\cite{Wollman1993,Tsuei1994,Kirtley2006} ( Fig.~\ref{fig:YBCO}(b)), which contrast with the single valued phase of $s$-wave superconductors,  has other important consequences.  This was early shown by  Tsuei~\emph{et al.}, who  built a tricrystal of yttrium barium copper oxide YBa$_2$Cu$_3$O$_{7-x}$ (YBCO) which naturally enclosed a phase gradient and thus showed spontaneous magnetization at low temperature~\cite{Tsuei1994},  and by the work of Il’ichev \emph{et al.} based on a YBCO bicrystal to realize exotic Josephson junctions with an anomalous magnetic field periodicity~\cite{Grajcar1999}. Indeed, the phase phase shift between $d$-wave lobes has strong implications when it comes to realizing $d$-wave based Josephson junctions and devices~\cite{Bauch2005,Ariando2006,Bauch2006,Cedergren2010}. One of the outstanding possibilities is the realization of junctions with a built-in $\pi$-shift between the electrodes, allowing for half-quantum superconducting loops~\cite{Hilgenkamp2003} that enable novel rapid single flux quantum logic devices architectures with enhanced capabilities~\cite{Ortlepp2006}. Furthermore, the combination of high-$T_\text{C}$ and the $\pi$ phase shift between $d$-wave lobes can be used for designing topological Josephson junction hosting Majorana bound states~\cite{Lucignano_2012,Lucignano_2013,Trani_2016}. Although extended to different $d$-wave/$s$-wave Josephson junctions ~\cite{Faley2020}, so far these $\pi$ junctions have not been explored using Dirac materials (and particularly graphene) as Josephson barrier. Yet, these devices would be endowed with an interesting property, namely tunability,  by replacing the usual passive weak link by a gate-tunable Dirac material. Graphene stems out as a material of choice in that prospect in that it is widely available, well characterized and has been implemented in low-temperature superconductor junctions.
Finally, another interesting property of all superconducting cuprates regarding superconducting proximity effects with Dirac materials is that their high-$T_\text{C}$ is accompanied by a superconducting gap whose amplitude can reach tens of meV, that is, one order of magnitude larger than that of conventional low-$T_\text{C}$. This is important because, as it will described below, many of the distinctive features of proximity effects in these materials are linked to the doping level, and particularly to the height of the Fermi level relative to the size of the superconducting gap. Yet, as discussed is Section~\ref{2D_material}, in real Dirac samples, including in single layer graphene, the doping (and thus the Fermi level) is often inhomogeneous at the nano-scale, with a variability that generally can reach  $\sim$~10 meV unless special, practically restrictive measures are taken ~\cite{dean2010boron}. Thus, one of the advantages of high-$T_\text{C}$ cuprates over conventional superconductors in the context of this review’s matter is that they offer a platform in which the possible electronic inhomogeneities of the graphene (and other 2D Dirac materials) can often be neglected when compared to the size of the superconducting gap.

Unfortunately, all of the interesting  properties  of cuprates are accompanied by a drawback: there are delicate materials for which the fabrication of solid-state devices to study proximity and Josephson effects in combination with 2D materials is often much more complicated than for conventional low-$T_\text{C}$ superconductors. The latter are usually metals (for example Al or Nb), whose growth techniques and conditions (\emph{e.g.} evaporation or sputtering at room temperature and inert atmosphere on many different substrates, epitaxial growth being unnecessary) make them compatible with standard nanofabrication approaches (\emph{e.g.} e-beam lithography combined with lift-off). This allowed the early fabrication of low-$T_\text{C}$ graphene-based Josephson devices~\cite{Heersche_2007}.  However, cuprates have a complex crystal structure and composition~\cite{Hott2016} --- see Fig.~\ref{fig:YBCO}(a) --- and the growth of thin films requires epitaxy on specific substrates at high temperatures (hundreds of degrees) in oxygen-rich atmosphere~\cite{Crassous2011,Swiecicki2012}. This makes them incompatible with the approaches traditionally used with low-$T_\text{C}$ metals.
Furthermore, the cuprate’s superconducting properties are extremely sensitive to the presence of structural defects~\cite{Tolpygo1996,Trastoy2014} and to the oxygen stoichiometry~\cite{Proust2019,Keimer2015a} ---which has made the demonstration of superconductivity in single layer BSCCO challenging ~\cite{Yu2019,Sandilands2010}, for example.  This sensitivity limits the conditions (temperature, chemicals) under which conventional nanofabrication approaches can be applied to combine $d$-wave cuprates with 2D materials. As discussed in Section~\ref{Device_Fabrication}, this has motivated the development of  \emph{ad hoc} fabrication techniques.

\section{Theory}\label{Theory}

\subsection{The normal/$d$-wave superconductor junction}\label{Theory_dwave_metal}
In this Section we provide a brief review on the  superconducting  proximity effect from a theory perspective.  It is well-know that metals and semiconductors in contact with superconductors acquire superconducting correlations due to the so-called proximity effect~\cite{Zagoskin_2014}.  The microscopic origin of such induced correlations is  the  process known as the Andreev reflection:  an incident electron forms a Cooper pair in the superconductor. Charge conservation at the interface results in the reflection of a hole~\cite{Zagoskin_2014}. In a ballistic system, the created electron-hole pair remains coherent over distances of the order of $v_\text{F}/E$, where $E$ is the energy of the electron measured from the Fermi-level, and $v_\text{F}$ the Fermi velocity. 
In metallic systems, the velocity of reflected holes is the opposite of one of the incoming electrons.  In  Dirac materials, like graphene and three-dimensional topological insulators, besides this so-called Andreev retro-reflection, Andreev reflection can be specular as well~\cite{Beenakker_2006,Beenakker_2008}. Manifestation of specular Andreev reflection  has been detected  experimentally in bilayer graphene~\cite{Efetov_2015}. The type of Andreev reflection can be changed from  retro to specular by  tuning  the electrochemical potential $\mu$ of the normal region: it is of specular type when $\mu$ is  close to the charge neutrality energy~\cite{Beenakker_2006}, and its of pure retro-type when $\mu$ is much larger than the superconducting gap energy~\cite{Linder_2007,Linder_2008}. In the following, we review the theoretical approach to Andreev reflection for both, metals and Dirac materials.  Importantly, throughout this  Section, we assume that the superconducting pair potential is spatially constant in the superconducting regions, and we ignore for simplicity the self-consistency of its spatial distribution. 

\vspace{0.25cm}

\begin{figure}[!t]
\centering
  \includegraphics[width=0.95\textwidth]{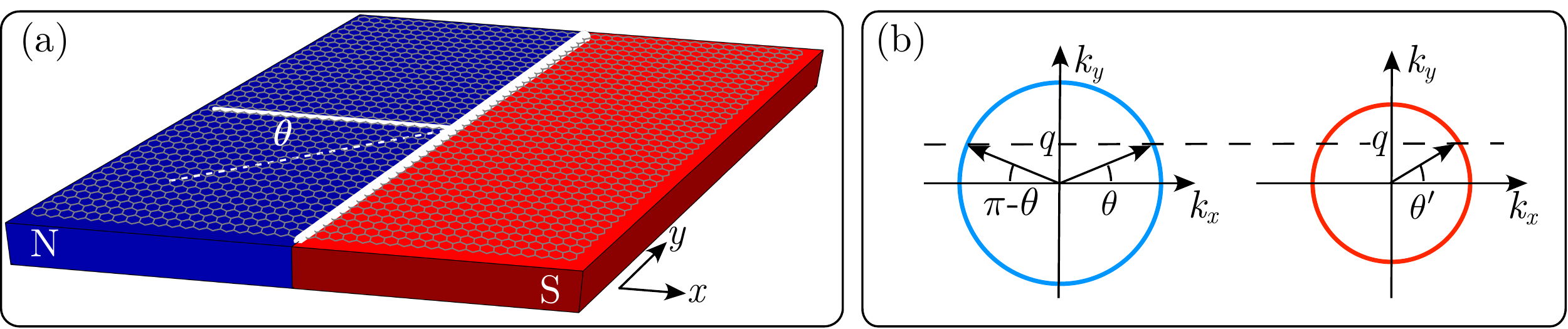}
  \caption{(a) Sketch of the N/S interface for the case of a parabolic metallic systems or for the case of  graphene. The thick white line separating the N/S region represent the BTK $Z$ parameter.  (b) Sketch of the angles relative to the normal reflection (blue circle) and  the Andreev reflection (red circle). The incoming carriers arrive with an angle $\theta$ with respect to the normal to the N/S interface, and are normal reflected with an angle $\pi-\theta$. Andreev reflected carriers propagate with an angle $\theta'$. }
  \label{fig:lateral}
\end{figure}

The Andreev reflection affects drastically  the transport properties of hybrid superconducting systems. 
In this review, we focus on lateral structures, {\it i.e.} systems on which superconducting electrodes are grown on top of a 2D or quasi-2D non-superconducting material (N), see Fig. \ref{fig:lateral}(a).   In such a setup,  the proximity effect in N leads to finite superconducting correlations with the same symmetry as the superconducting lead.  The main transport features are then obtained by analyzing a planar hybrid structure between superconducting (those proximitized by the superconducting leads) and normal regions --- Fig. \ref{fig:lateral}(a). 
%
%
%
%
We focus here on clean ballistic systems and introduce the   Bogoliubov-de Gennes (BdG) equation
%
%
\begin{subequations}\label{BdG}
\begin{align}\label{BdG_equation}
    \mathcal{H}_\text{BdG}\circ{\bf \Psi}=E{\bf \Psi}\;, 
\end{align}
%
%
where
%
%
\begin{align}\label{BdG_Hamiltonian}
    \mathcal{H}_\text{BdG}(\bm{r}, \bm{r'})=\begin{pmatrix}
[H(\bm{r})-\mu(\bm{r})]\delta (\bm{r}-\bm{r'})& \Delta(\bm{r},\bm{r'}) \\
\Delta ^*(\bm{r},\bm{r'})& -[\bar H(\bm{r})-\mu(\bm{r})]\delta (\bm{r}-\bm{r'})
\end{pmatrix}\;.
\end{align}
\end{subequations}
%
%
and the $\circ$ in Eq. (\ref{BdG_equation}) denotes convolution in real space. Because our focus is on $d$-wave superconductors, we need to keep the two components dependence in the effective superconducting pairing potential induced in the N region, $\Delta$~\cite{bruder1990andreev}.  The above  Hamiltonian is a matrix in the  particle-hole space. Each of its components are in the general also  matrices whose structure depends on the specific material. The term $H$ describes the  normal state Hamiltonian, whereas $\bar H=\mathcal{T}H\mathcal{T}^{-1}$ is its time-conjugated, here $\mathcal{T}$ denotes the time-reversal symmetry (TRS) operator. In a hybrid structure $\Delta$ and the chemical potential $\mu$ in Fig.~\ref{fig:lateral}(a) are step-like functions.   

Because of the anistropy in $\Delta$, the  search for an exact solution of  Eqs.~\eqref{BdG} is a difficult  task. 
Indeed, the spinors  $\bm{\Psi}$ oscillates over a scale of the inverse  Fermi wave vector $k_\text{F}^{-1}$, whereas the superconducting correlations usually vary over a  much longer scale $\xi_\text{S}$. In this case one can   single out the fast oscillations by the substitution $\bm{\Psi}\rightarrow \text{e}^{-\text{i}\bm{k}_\text{F}\cdot\bm{r}}\bm{\Psi}$. In leading order in $k_\text{F}^{-1}$, Eqs.~\eqref{BdG} reduce to the well-known Andreev equations~\cite{bruder1990andreev,tanaka1995theory}:
%
%
\begin{equation}
    \begin{pmatrix}
-\text{i}\frac{\bm{k}_\text{F}}{m}\cdot\bm{\nabla}& \Delta(\bm{k}_\text{F},\bm{R}) \\
\Delta ^*(\bm{k}_\text{F},\bm{R})& \text{i}\frac{\bm{k}_\text{F}}{m}\cdot\bm{\nabla}
\end{pmatrix}
{\bf \Psi}=E{\bf \Psi}
\;, 
\label{AndreevEq}
\end{equation}
%
%
where  $\Delta(\bm{k}_\text{F},\bm{R})$ is the Wigner transform of  $\Delta(\bm{r},\bm{r'})$ evaluated at the Fermi level, and $\bm{R}$ is the center of mass coordinate. 

For the geometry shown in Fig.~\ref{fig:lateral}(a), the N/S interface is located at $x=0$. The pair potential is only finite at $x>0$.
If we denote with $\bm{k}_\pm$ the incident and reflected wavevectors then the pair potential of a $d_{x^2-y^2}$ is given by   $\Delta(\bm{k}_\pm)=\Delta_0\cos\left[2(\theta\mp \alpha)\right]$, where $\theta$ is the incident angle of an electron from the normal conductor, and $\alpha$ the angle between the normal to the interface and the crystalline axis along which the pair potential reaches its maximum value.  As in the original work by Blonder Tinkham Klapwijk (BTK)~\cite{Blonder_1982}, an imperfect N/S barrier is modeled by a delta-like potential, $U(x)=\mathcal{U}_0\delta(x)$. 
%
%
\begin{figure}[!t]
\centering
  \includegraphics[width=0.95\textwidth]{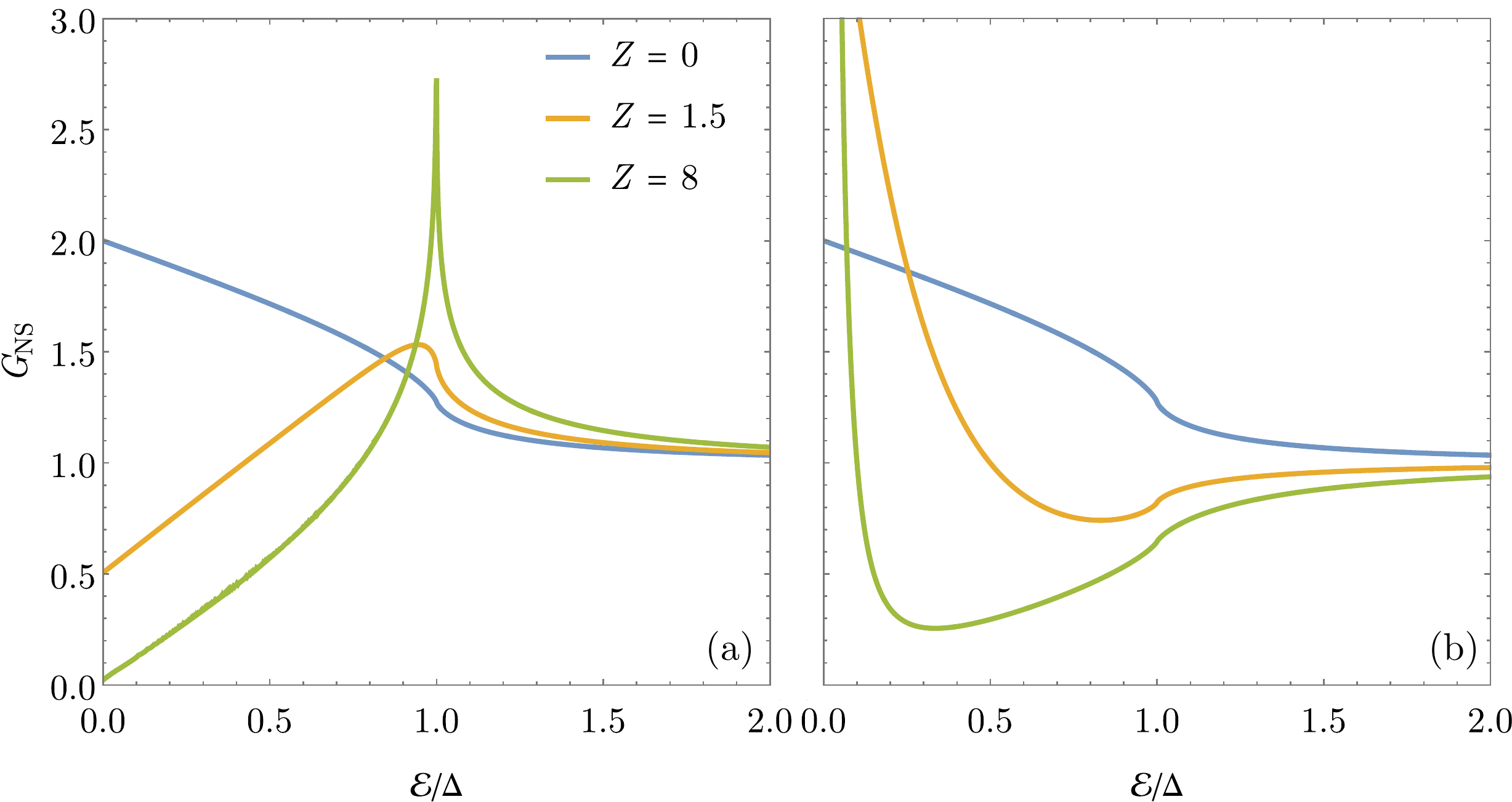}
  \caption{The normalized conductance $G_\text{NS}$ as a function of the energy of injected electrons for (a) $\alpha=0$ and (b) $\alpha=\pi/4$. For both panels, we have used three different values of the $Z$ parameter~\cite{tanaka1995theory}.\label{fig:dwave_normal}}
  \end{figure}
 %
 %
This term leads, to complement the continuity of the wave function with the following boundary condition for Eq.~\eqref{AndreevEq}:
%
%
\begin{align}\label{BoundCond}
\partial_x{\bf \Psi}^\text{S}(0^+)-\partial_x{\bf \Psi}^\text{N}(0^-)=2m\mathcal{U}_0{\bf \Psi}(0)\;.
\end{align}
%
%
The boundary problem Eqs.~\eqref{AndreevEq} and \eqref{BoundCond} can be solved straightforwardly. The electronic transport properties through the  N/S interface are determined by the Andreev, $r_\text{A}$,  and normal, $r_\text{N}$,  reflection amplitudes, of the electrons and hole injected from the normal metal. From the BTK theory the conductance at low temperatures can be written as \cite{Chaudhuri_1995,Mortensen_1999}:
%
%
\begin{equation}
    \label{conductanceBTK}
    G_\text{NS}(V)=\frac{e^2 N_\text{F}v_\text{F}w}{\pi}\int_{-\pi/2}^{-\pi/2}d\theta\left( 1+|r_\text{A}(V,\theta)|^2-|r_\text{N}(V,\theta)|^2\right)\; ,
\end{equation}
%
%
where $N_\text{F}$ is the (2D) density of states at the Fermi level and $w$ the width of the junction. 
The conductance of a N/$d$-wave superconductor  was first discussed and calculated in Refs.~\cite{Hu_1994,tanaka1995theory}. 
The main results of those theories are summarized  in Fig.~\ref{fig:dwave_normal} where we show the dependence of the normalized conductance ($G_\text{NS}=G_\text{S}/G_\text{N}$ with respect to the normal value $G_\text{N}$) through the N/S interface as function of the injection energy (applied voltage). Different curves correspond to different values of the BTK parameter $Z=2m\mathcal{U}_0/k_\text{F}$. In the left panel the angle $\alpha$ between the normal to the interface  and the direction along which the gap is maximum, is zero. In this case $\Delta(\bm{k}_+)=\Delta(\bm{k}_-)=\Delta_0$, and the behaviour of the conductance resembles the one of an isotropic N/S junction. The zero-bias conductance reaches its maximum, $G_\text{NS}(0)=2$ value when the interface is transparent ($Z=0$). For a low-barrier transmission, $Z\gg1$, the conductance approaches the  tunneling behaviour. 
If the interface is transparent then the  conductance do not depend on the angle $\alpha$ (blue solid lines in both panels of Fig.~\ref{fig:dwave_normal}).
The situation changes drastically when both  $\alpha\neq0$ and $Z$ are finite, as shown in the right panel of Fig.~\ref{fig:dwave_normal}. In this case  the zero bias conductance  may largely exceed the value of $2$ and it diverges as $G_\text{NS}(0)\sim  Z^2$ when $Z\rightarrow\infty$. This  is due to the fact that, whereas  the normal conductance, $G_\text{N}$ is monotonically reduced by increasing $Z$, $G_\text{S}$ remains equal to $2$ for certain values $E_0$ of the energy where resonance tunneling through bound-states takes place. Such bound-states are formed, in analogy to conventional S-S ballistic junctions,  because normal  reflected quasiparticles from the superconductor experiences pair-potentials with different phases, before and after the reflection~\cite{Kashiwaya_1996,Kashiwaya2000}.

\subsection{Proximity effects $s$-wave/Dirac material: Specular Andreev reflection}\label{sec_s_graphene}

\subsubsection{Graphene junctions}\label{sec_s_graphene2}
%
%
\begin{figure}[!h]
\centering
  \includegraphics[width=0.95\textwidth]{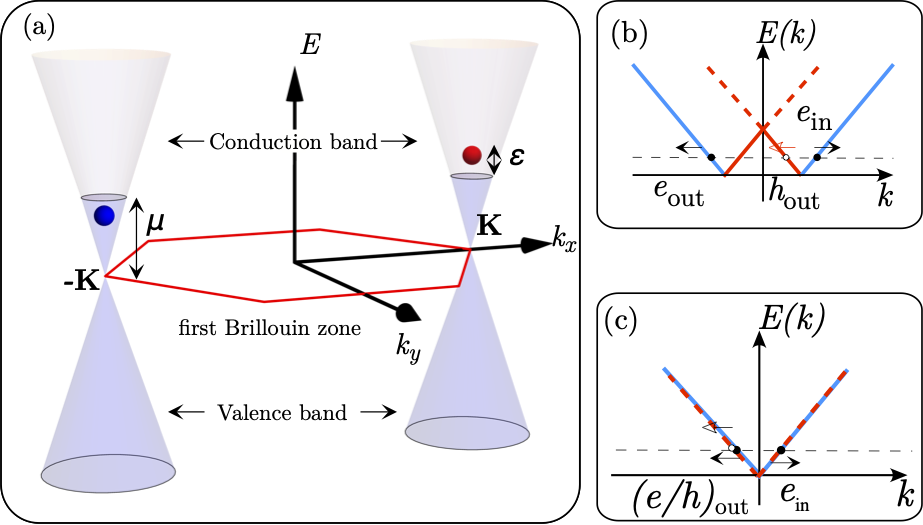}
  \caption{\label{fig_Dario_One}(Color online) (a) Electron and hole excitations in the conical band structure of system for electrons with graphene energy dispersion, red and blue sphere at energy states at $\mu\pm\epsilon$ are converted into each other by Andreev reflection at the normal/superconductor interface.  Panel (b) and Panel (c): Sketch of the excitation  spectrum for electron in the conduction band (blue solid line) and holes in the conduction band (red solid line) and valence band (red dashed line) for the case in which $\mu\gg\text{Max}[\Delta,\epsilon]$ and $\mu\ll\text{Max}[\Delta,\epsilon]$, respectively~\cite{Beenakker_2008,Bercioux_2018}.}
\end{figure}
%
%
In this Section, we  show how the process of Andreev reflection changes when considering a Dirac-like Hamiltonian in the BdG Eq.~\eqref{BdG_Hamiltonian}. First, we  consider the case of graphene proximitized with a $s$-wave superconductor~\cite{Beenakker_2006}.  In the Section~\ref{Theory_dwave_graphene} we  generalize the results to the case of a $d$-wave superconductor. 


As mentioned in the introduction, the energy spectrum of graphene is almost linear  around the two Dirac  points, $\mathbf{K}$ and $\mathbf{K}'$,  of the first Brillouin zone~\cite{Wallace_1947}, see Fig.~\ref{fig_Dario_One}. Here, we can perform the so-called low-energy long-wavelength approximation (LWA), and  introduce the following effective Hamiltonian describing the electron behaviour around Dirac points~\cite{Goerbig_2011}:
%
%
\begin{align}\label{ham_lwa}
    \mathcal{H}_\text{LWA} = v_\text{F} \tau_0(\sigma_x p_x +\sigma_y p_y)
\end{align}
%
%
where $v_\text{F}\sim 10^6$~m/s is the effective Fermi velocity of the electron, $\bm{p}$ is the electron momentum operator, $\tau_0=\mathbb{I}_2$ represents the identity matrix associated to the two  Dirac cones,  and  $\bm{\sigma}=(\sigma_x,\sigma_y)$ are the Pauli matrices associated to the so-called pseudo-spin connected to the two carbon atoms A and B in the unit cell of graphene, see Fig.~\ref{fig_one}. We  disregard the fermionic spin degree-of-freedom since we only consider situations in which  time-reversal symmetry (TRS) is preserved. The Hamiltonian in Eq.~\eqref{ham_lwa} acts on  the following  spinor
%
%
\begin{equation*}
\Psi=(\psi_\text{A},\psi_\text{B},-\psi_\text{B}',\psi_\text{A}'),   
\end{equation*}
%
%
where the first two components are associated to the valley $\mathbf{K}$ and the last two to $\mathbf{K}'$, and the indices A and B are associated to the pseudo-spin. In order to construct the BdG Hamiltonian~\eqref{BdG_Hamiltonian},  for proximitized graphene it is important to notice that 
according to Eq.~\eqref{ham_lwa}  the electrons in  valleys $\mathbf{K}$ and $\mathbf{K}'$ are related to each other via TRS~\cite{Beenakker_2008}. When considering graphene in the LWA, this symmetry operator can be expressed as $\mathcal{T}=-(\tau_y\otimes\sigma_y)\mathcal{K}$,
where $\tau_y(\sigma_y)$ are the $y$ Pauli matrices for the valley(pseudo-spin) degree-of-freedom, and $\mathcal{K}$ is the complex conjugation operator~\cite{Haake_2010}. In other words, within  the LWA, time-reversed carriers are associated to the two different valleys.  When constructing the BdG matrix Hamiltonian, we can then split the $8\times8$ problem into two independent  $4\times4$ problems.  Importantly, we assume that  all the spatial modulation involved in the problem takes place on a length scale much larger than the graphene lattice spacing $a_0$. 
This results in the following BdG Hamiltonian for graphene, 
%
%
\begin{align}\label{BdG_graphene}
\mathcal{H}_\text{BdG}^\text{LWA}= \begin{pmatrix}
v_\text{F} \bm{\sigma}\cdot\bm{p}-\mu & \Delta(\bm{r}) \\
\Delta(\bm{r})^* & \mu-v_\text{F} \bm{\sigma}\cdot\bm{p}
\end{pmatrix}.
\end{align}
%
%
Within this model  the pairing potential does not mix the component of the spinors, thus $\Delta(\bm{r})=\Delta_0 \sigma_0$, this correspond to consider only intraband superconducting pairing on each sublattice A and B~\cite{footnoteLWA}. 

The chiral semimetallic nature of graphene enriches  the physics of the electron scattering at a N/S interface. An electron from the conduction band impinging on the interface with an angle $\theta$ is normal reflected within the same band with opposite angle $\pi-\theta$ or can be Andreev reflected as hole either in the conduction or in the valence band. The former process is a retro-reflection whereas the latter is a  specular-reflection~\cite{Beenakker_2006}.

In the practice  we are dealing  with  a genuine two-dimensional scattering problem with two free parameters, the injection energy $\epsilon$ and the momentum parallel to the N/S interface $q$ --- see Fig.~\ref{fig:lateral}(b). Considering an incoming electron in the conduction band, the incidence angle $\theta$ in polar coordinate reads:
%
%
\begin{align}\label{injection:angle}
\theta& = \arcsin\left[\frac{\hbar v_\text{F} q}{\epsilon+\mu}\right],
\end{align}
%
%
where $(\epsilon+\mu)(v_\text{F} \hbar)^{-1}=|\bm{k}|$ is the modulus of the momentum for the conduction band.
For a hole, the value of the reflection angle can be obtained by simple kinematic considerations: conservation of the two free parameters in the scattering process --- see Fig.~\ref{fig:lateral}(b). Thus the angle of propagation of the hole is
%
%
\begin{align}\label{reflection:angle}
\theta'& = \arcsin\left[\frac{\hbar v_\text{F} q}{\epsilon-\mu}\right].
\end{align}
%
%
We can also treat along the same line of reasoning  the case of a hole  injected state. The propagation direction of the hole strongly depends on the value of the injection energy compared to the chemical potential.
In Fig.~\ref{fig:lateral}(b) we show a sketch of the the Fermi circles for electrons and holes in the normal region; if the momentum $q$ exceeds the radius of the Fermi circle for holes, the corresponding Andreev reflected state is suppressed. Thus, we introduce a critical injection angle defined as:
%
%
\begin{equation}\label{alphacritical}
\theta_\text{c} =\arcsin\left[\frac{|\epsilon-\mu|}{\epsilon+\mu}\right]\,. 
\end{equation}
%
%

When treating this scattering process, the main difference compared to the case of electrons in metals arises in the boundary conditions, c.f. Eq.~\eqref{BoundCond}. 
For  graphene it is sufficient to require the continuity of the wave function at the interface. We can solve this system of equations analytically and obtain an expression for the normal $r_\text{N}$ and Andreev $r_\text{A}$ reflection~\cite{Beenakker_2006,Bercioux_2018}. These reflection amplitudes read:
%
%
\begin{subequations}\label{normal_andreev_exact}
\begin{align}
    r_\text{A} &= \begin{cases} 
\text{e}^{\frac{\text{i}}{2}(\theta+\theta'-\pi)}\Xi^{-1}\sqrt{\cos(\theta)\cos(\theta')} & \text{if}~|\theta|<\theta_\text{c}\\
0 & \text{if}~|\theta| > \theta_\text{c}
\end{cases} \,, \\
r_\text{N} & = \text{e}^{\frac{\text{i}}{2} (2\theta-\pi)} \Xi^{-1} \left[\cos\beta \sin\left(\frac{\theta+\theta'}{2}\right) + \text{i} \sin\beta\sin\left(\frac{\theta-\theta'}{2}\right)\right]\,, \\
\Xi& = \cos\left(\frac{\theta+\theta'}{2}\right)\cos\beta+\text{i} \cos\left(\frac{\theta+\theta'}{2}\right)\sin\beta\; ,
\end{align}
\end{subequations}
%
%
where  we have introduced the function $\beta(\epsilon)$ defined as
%
%
\begin{align}\label{beta}
\beta(\epsilon)=\begin{cases}
\arccos\left(\frac{\epsilon}{\Delta}\right) & \epsilon<\Delta \\
-\text{i}\, \text{arccosh}\left(\frac{\epsilon}{\Delta}\right) & \epsilon>\Delta
\end{cases}\,.
\end{align}
%
%
It is important to note that in the previous derivation, following Ref.~\cite{Beenakker_2006}, we have performed the so-called Andreev approximation~\cite{Zagoskin_2014} only in the superconducting region, where  we assume that  the local chemical potential $\mu_\text{S}$ is the dominant energy scale: $\mu_\text{S}\gg\text{Max}[\Delta,\epsilon]$.

We analyse now the amplitudes in Eqs.~\eqref{normal_andreev_exact} similarly to Ref.~\cite{Beenakker_2006}.  The regime of Andreev retro- or specular-reflection can be addressed by considering the chemical potential in the normal region $\mu$. If the chemical potential is the dominating energy scale $\mu\gg\text{Max}[\Delta,\epsilon]$, then the hole generated by an Andreev refection results in an empty states in the conduction band with  a propagation angle  opposite  to the one of the  incoming electron $\theta'=-\theta$. This is the limit of Andreev retro-reflection, here we can write a simplified expression for the 
reflection amplitudes [c.f. Fig.~\ref{fig_Dario_One}(b)]:

%
%
\begin{subequations}\label{retro}
\begin{align}
r_\text{A}& =\frac{\text{i}\cos\theta}{\zeta + \frac{\epsilon}{\Delta}\cos\theta}\,,\label{AR} \\
r_\text{N}& = \frac{\text{i}\text{e}^{\text{i} \theta}\zeta\sin\theta}{\zeta + \frac{\epsilon}{\Delta}\cos\theta}\,. \label{NR}
\end{align}
\end{subequations}
%
%
In the opposite regime, $\mu\ll\text{Max}[\Delta,\epsilon]$, the hole generated by  the Andreev reflection is in the valence band and  $\theta'=\theta$ --- this is the case of Andreev specular-reflection~\cite{Beenakker_2006}. Also in this case, we can simplify  Eqs.~\eqref{normal_andreev_exact} for the normal and Andreev reflection amplitudes [c.f. Fig.~\ref{fig_Dario_One}(c)]:
%
%
\begin{figure}[!t]
    \centering
    \includegraphics[width=0.95\textwidth]{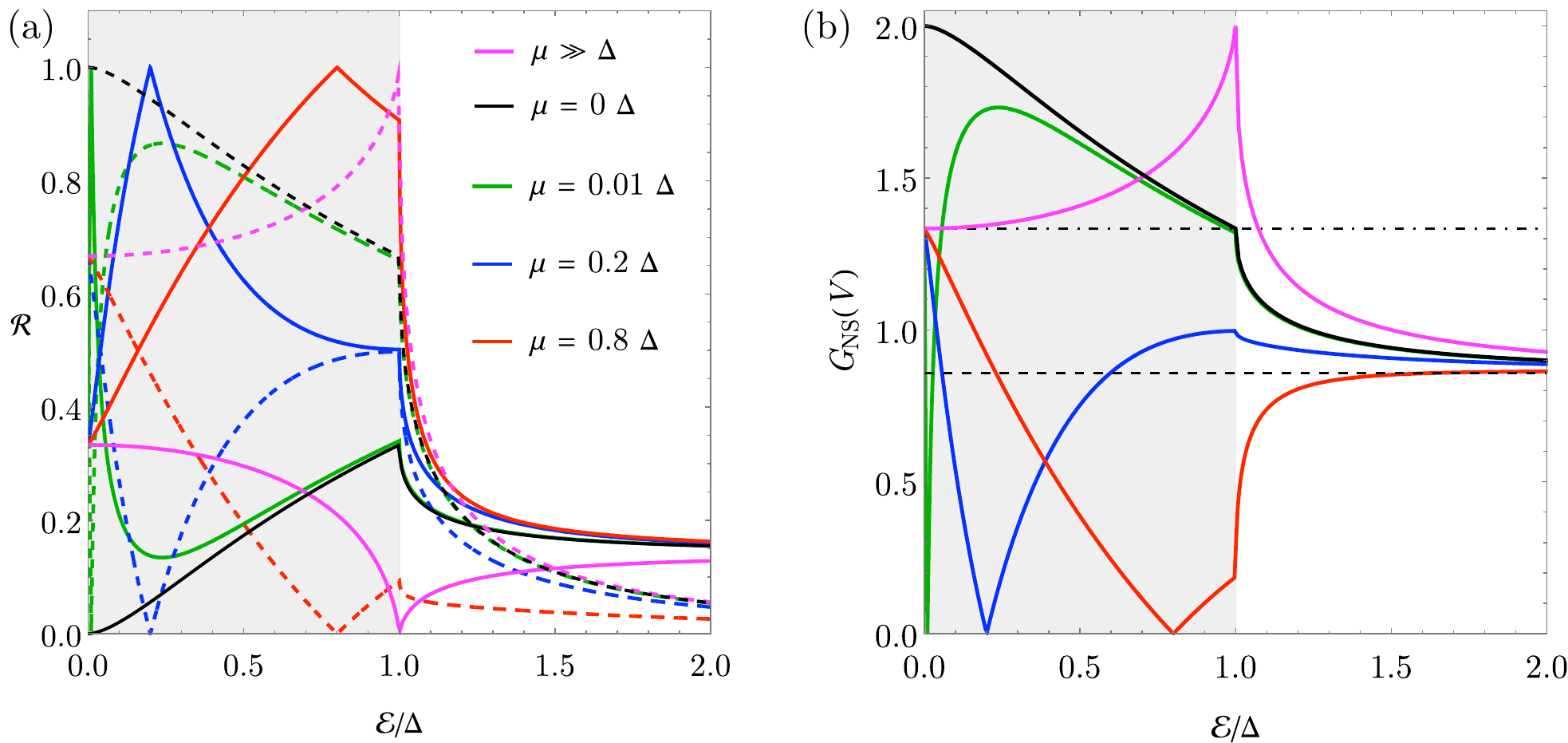}
    \caption{(a) Normal (solid-line) and Andreev (dashed-line) reflection as a function of the injection energy $\epsilon$, for various values of the chemical potential $\mu$. (b) Differential conductance as a function of the injection energy $\epsilon$ for the same values of the chemical potential as in (a). In panel (b), the dotted-dashed line corresponds to the differential conductance at zero applied voltage, whereas the dashed line represents the conductance of the normal state. In both panels, the grey area represents injection energy smaller than the superconducting energy gap $\Delta$, this represents the scale of references for the energies. The results in the two panels are in agreement with Refs.~\cite{Beenakker_2006,Bercioux_2018}.}
    \label{fig_ref_cond}
\end{figure}
%
%
%
%
\begin{subequations}\label{specular}
\begin{align}
r_\text{A}& =\frac{\text{i}\text{e}^{\text{i}\theta}\cos\theta}{\frac{\epsilon}{\Delta} +\zeta \cos\theta}\,,\label{AS} \\
r_\text{N}& = \frac{\text{i}\text{e}^{\text{i} \theta}\frac{\epsilon}{\Delta}\sin\theta}{\frac{\epsilon}{\Delta} +\zeta \cos\theta}\,. \label{NS}
\end{align}
\end{subequations}
%
%
In Eqs.~\eqref{retro} and~\eqref{specular}  we have introduced the following  function:
%
%
\begin{align}\label{zeta}
\zeta=\begin{cases}
\text{i} \sqrt{1-\left(\frac{\epsilon}{\Delta}\right)^2} & \epsilon < \Delta \\
\sqrt{\left(\frac{\epsilon}{\Delta}\right)^2-1} & \epsilon>\Delta
\end{cases}\,.
\end{align}
%
%

In Fig.~\ref{fig_ref_cond}(a), we present the normal and the Andreev reflection as a function of the injection energy and for various values of the chemical potential of the normal region $\mu$ to show the cross-over from specular to retro Andreev reflection. For injection energies $\epsilon$ smaller that the superconducting gap $\Delta$, we observe that the Andreev reflection goes to zero when the injection energy is equal to $\mu$, whereas at these energies the normal reflection equals one. For injection energies larger than the superconducting gap, the normal reflection is finite whereas the Andreev reflection goes to zero by increasing the energy. 

The differential conductance for the case of graphene is obtained by Eq.~\eqref{conductanceBTK}, using $N_\text{F}=\frac{W}{\pi}\frac{\mu+\epsilon}{\hbar v_\text{F}}$ for the graphene density of states, where $W$ corresponds to the finite transverse dimension of the graphene sample.
As a consequence of the crossover between specular and retro-reflection of the Andreev channel, the differential conductance of the N/S interface is zero when the injection energy coincides with the chemical potential of the normal region $\mu$, this is shown in Fig.~\ref{fig_ref_cond}(b). An additional consequence of the graphene linear dispersion and of having performed the Andreev approximation only in the superconducting region is in the value of the differential conductance at zero bias. This is independent of the chemical potential and equal to $4/3G_0(eV)$, where $G_0(eV)=2e^2/h N_\text{F}$. Another interesting limit studied in Ref.~\cite{Beenakker_2006} is  the limit of large applied bias where, for any chemical potential, we find the universal value of the differential conductance of $(4-\pi)G_0(eV)$.  These limiting values are  consequence of having imposed the Andreev approximation only in the superconducting region. However, if this condition is relaxed, the limiting value of the normal conductance goes back to $G_0$, whereas the zero bias conductance equals  $2G_0$ as  shown in Refs.~\cite{Linder_2007,Linder_2008}. 

\subsubsection{3DTI junctions}

The results obtained so far are not bounded to graphene as a Dirac material but are strongly connected to its 2D linear dispersing kinetic term. Therefore, they can be extended to the case of 3DTIs proximitized by a $s$-wave superconductor. For these systems, the main difference arises in the structure of the superconducting pairing potential in Eq.~\eqref{BdG_graphene}. This is due to a different basis describing the spinfull Hamiltonian of a 3DTI, that in this case reads $\Psi=(\psi_\uparrow,\psi_\downarrow,\psi_\uparrow^\dag,\psi_\downarrow^\dag)$~\cite{Tkachov_2013_a,Tkachov_2013_b}.  The pairing potential for the case of a proximitized 3DTI is 
%
%
\begin{equation}\label{pairing_3DTI}
    \hat{\Delta}=\begin{pmatrix}
        0 & \text{i} \sigma_2 \Delta(\bm{r}) \\
        -\text{i} \sigma_2\Delta^*(\bm{r}) & 0
    \end{pmatrix}\,.    
\end{equation}
%
%
Utilizing this pairing potential and the kinetic term of a 3DTI, the results for the normal and Andreev reflection are identical to the ones described by  Eqs.~\eqref{normal_andreev_exact}. As a consequence, the physics of the Andreev specular- and retro-reflection also appears  in 3DTIs~\cite{Majidi_2016,Bercioux_2018}. Importantly, as a consequence of the non-trivial spin structure associated to the spin-momentum locking, the proxi\-mitized structure behaves as an effective $p$-wave superconductor when the proximitizing superconductor is of $s$-wave type~\cite{Gorkov_2001}. In this case  the spin pairing has a triplet structure. A magnetic field can modify the nature of the proximity effect by mixing singlet pairing arising from the proximitizing $s$-wave superconductor and the $p$-wave from the 3DTI. This mixing will result in  different differential conductance of hybrid structure of 3DTI/F/S, where F denotes a ferromagnet with an exchange field or a  Zeeman field associated to the magnetic field~\cite{Burset_2015}.
Triplet order parameter has been suggested for many applications including the  proposal for a perfect spin-triplet filter~\cite{Breunig_2018}.
This type of heterostructure can host  zero-energy Majorana bound states~\cite{Qi_2009,Fu2008,Tanaka_2009,Law_2009}. 
Hybrid 3DTI/$s$-wave exhibit a cooling power that is also depending on the two possible regime of the Andreev reflection, larger for specular and smaller for retro-reflection~\cite{Bercioux_2018}.



\subsubsection{Interface transparency}

When considering the hybrid junction between a normal metal and a superconductor, we have seen that it is possible to model the transparency of the interface via a single parameter $Z$~\cite{Blonder_1982}. On the contrary, a similar procedure is not permitted in a Dirac system, since the only boundary condition is the continuity of the wave function at the interface~\cite{Beenakker_2008,Katsnelson_2006,Young_2009,Tudorovskiy_2012}. In  non-Dirac systems, the main role played by a finite $Z$-parameter is to increase the  magnitude of normal reflection channel at  the N/S interface. We see from the results in Fig.~\ref{fig_ref_cond}(a), that in a two-dimensional Dirac system, a channel of normal reflection is intrinsically present and that it can be modulated by tuning the chemical potential of the normal region. Thus, a strategy to \emph{mimic} a $Z$-parameter in a Dirac system consists in considering a finite size electrostatic barrier of width $w$ and height $\mathcal{U}$. Similarly to the case of charge carriers with parabolic dispersion, we can obtain the limiting case of a delta-like barrier  by sending $w\to0$ and $\mathcal{U}\to\infty$ keeping finite the ratio $\mathcal{U}w/(\hbar v_\text{F})\equiv Z$. This approach has used  in Refs.~\cite{Bhattacharjee_2006,Linder_2007,Linder_2008}, and the approach can be recast into a boundary problem for scattering matrices~\cite{HLi_2017}. Following this approach,  a periodic oscillation of differential conductance as a function of the injection energy was found, this result is entirely associated to the Dirac-like dispersion. The period of the oscillations is proportional to the parameter $Z$~\cite{Linder_2007,Linder_2008,Bhattacharjee_2006}. When graphene is paired with a $s$-wave superconductor, the differential conductance evaluated this way and with equal chemical potential in the normal and superconducting region, resemble the one obtained for electrons with a quadratic kinetic term, however, for large $Z$ it is not possible to find a complete suppression of the differential conductance~\cite{Linder_2008}, as a consequence of the  Klein tunneling taking  place at the interface~\cite{Linder_2008,Katsnelson_2006}.

\subsection{Proximity effect $d$-wave/Dirac material}\label{Theory_dwave_graphene}

%
%
\begin{figure}[!ht]
    \centering
    \includegraphics[width=0.95\textwidth]{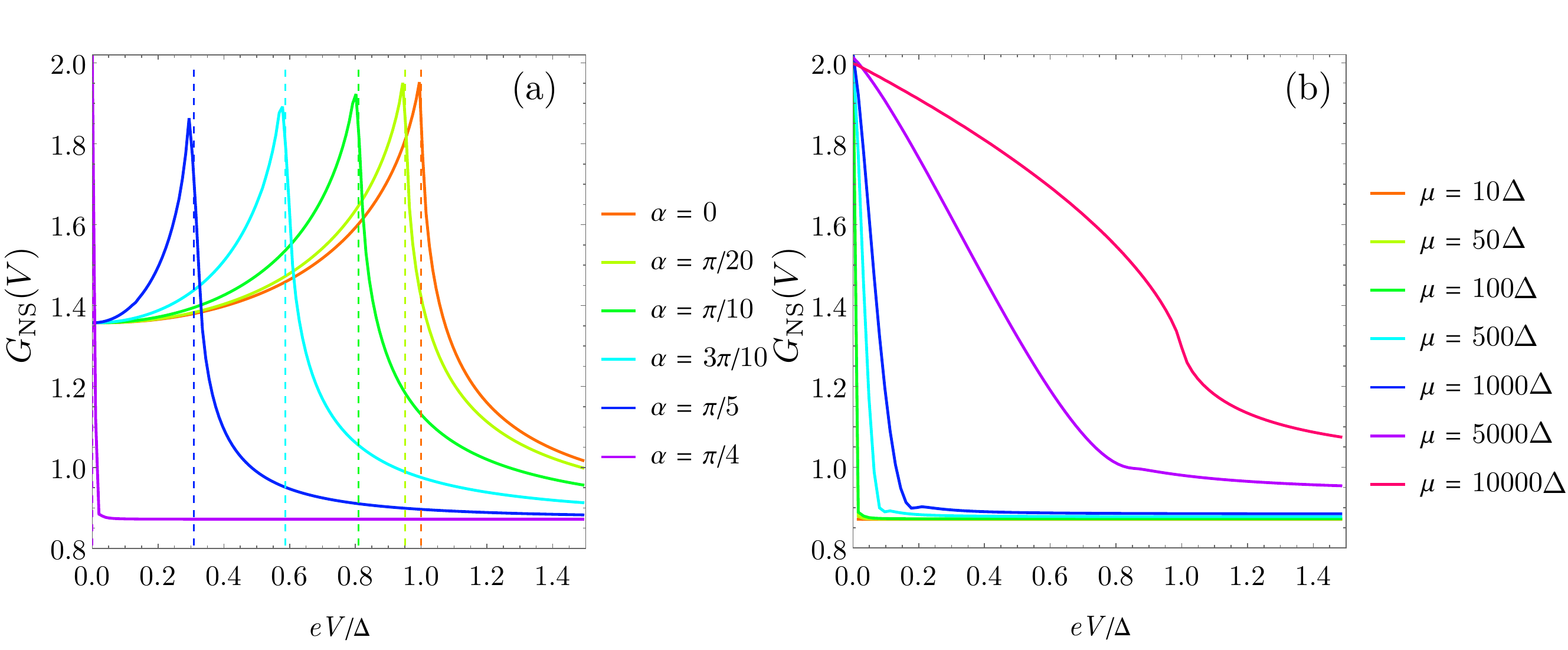}
    \caption{\label{fig:$d$-wave}(a) Differential conductance of the hybrid graphene/$d$-wave superconductor interface as a function of the applied voltage for different orientations of the superconducting order parameter $\alpha$. The vertical dashed lines represents the value of the reduced gap for the corresponding value of $\alpha$ accordingly to the expression $\Delta(\alpha)=\cos(2\alpha)$. The peaks of the differential conductance coincide with the gap value only for large values of $\mu_\text{S}$, otherwise, the peaks gets smoother and depart from the value of the gap~\cite{Linder_2007,Linder_2008}. (b) Zero-bias conductance peak at $\alpha=\pi/4$ as a function of the applied voltage for various values of the chemical potential $\mu$ in the normal region. In both panels we have set the chemical potential of the superconducting region to $\mu_\text{S}=10^4\Delta$. The results in the two panels are in agreement with Refs.~\cite{Linder_2007,Linder_2008}.}

\end{figure}
%
%
\subsubsection{Graphene junctions}

We consider a  $d_{x^2-y^2}$-wave superconductor with an order parameter  described by the function $\Delta(\bm{k})=\Delta_0\cos[2(\theta-\alpha)]$, where $\theta=\arctan(k_y/k_x)$ parametrize the injection direction, and $\alpha$ describes different orientation of the gap in the $\bm{k}$-space with respect to interface~\cite{Linder_2007,Linder_2008} --- see Fig.~\ref{fig:YBCO}(b). The differential conductance can be obtained following the same procedure we used for the case of $s$-wave pairing, but accounting for a momentum dependent order parameter~\cite{Linder_2008}. This can change in magnitude with $\alpha$ being larger for $\alpha=0$ and zero for $\alpha=\pi/4$ and with opposite signs of the different lobes~\cite{Kashiwaya_1995,Hu_1998}. We assume, as we did for the case of proximity with a $s$-wave superconductor, that the order parameter induced in graphene is of intraband type, \emph{i.e.} pairing sites of the same sublattice~\cite{Beenakker_2006,Linder_2007}. The behaviour of the hybrid N/S junction for a graphene proximitized with a $d$-wave superconductor, differs considerably compared to the case of electron with a quadratic kinetic term~\cite{Linder_2007,Linder_2008} that we considered at the beginning of this Section. Here we summarize the main differences and the possible causes:
\begin{itemize}
    \item The zero-bias conductance peak (ZBCP) is observed in the graphene/$d$-wave hybrid junction only if $\alpha=\pi/4$ and in the case of neglecting the Andreev-approximation in the normal part of the junction. This can be explained by the fact that only for finite $\mu$ in the normal region and for $\mu\neq\mu_\text{S}$, it is possible to achieve a finite channel for normal reflection. This channel is essential to form the zero energy bound state~\cite{tanaka1995theory,Hu_1994}.
    \item For angles different from $\alpha=\pi/4$, the peak sits to a different energy, reflecting the reduction of the energy gap expressed by the function $\Delta(\bm{k})$. This behaviour depends on the strength of the chemical potential in the superconducting region, the peaks appear exactly at the position of the gap for large $\mu_\text{S}$, whereas they get smoother and have a maximum for a lower energy compared to the gap for smaller $\mu_\text{S}$~\cite{Linder_2007,Linder_2008}.
    \item Contrary to the case of quadratic kinetic terms, the ZBCP does not diverge because the conductance of the normal state $G(\epsilon\to0)\neq0$, therefore the ZBCP is always bound to $2G_0$.  
\end{itemize}
These features are summarized in the two panels of  Fig.~\ref{fig:$d$-wave}, where the differential conductance $G_\text{NS}(V)$ is calculated using Eq.~\eqref{conductanceBTK}.
For equal chemical potential in the normal and superconducting graphene regions, we recover a result similar to the one obtained for electrons with quadratic dispersion when $Z=0$~\cite{tanaka1995theory,Linder_2007,Linder_2008} --- see Fig.~\ref{fig:dwave_normal}. This is shown in Fig.~\ref{fig:$d$-wave}(b) for the case $\mu\equiv\mu_\text{S}$.

We note in passing that the assumption of intraband constraint for the proximity effect can be relaxed. It was proven that choosing a pairing beyond the intraband constraint can lead to a $d$-wave pairing in intrinsic graphene~\cite{Black_Schaffer_2007,Black-Schaffer_2014}. Additionally, self-consistent calculations have shown that the induced pairing by a $d$-wave superconductor leads to an effective $p_x$- or $p_y$-wave pairing symmetry depending on the orientation of the boundary, \emph{i.e.} armchair or zigzag~\cite{Linder_2009}. However, the authors note that this effective $p+\text{i} p$ wave pairing does not break neither TRS nor produce spin-triplet states~\cite{Linder_2009}. 

\subsubsection{3DTI junctions}
In the case of a 3DTI as Dirac material, we have to include the momentum dependent order parameter as we have done in Eq.~\eqref{pairing_3DTI}. Here we account for the anisotropy of the order parameter with $\Delta\to\Delta(\bm{k})$~\cite{Linder_2010}. Interestingly, because of the spin-momentum locking, the ZBCP we have seen for the case of graphene or in a $d$-wave superconductor, moves from being two-fold degenerate to non-degenerate. As a consequence, this state describes a zero energy Majorana fermion with peculiar dependence on external magnetization~\cite{Linder_2010,Sato_2009,Zhang_2013}. Here, it was shown that the ZBCP splits into two separate peaks by introducing a Zeeman field perpendicular to the N/S interface, this behaviour is opposite to what expected for the case of proximization with a $s$-wave superconductor. Additionally, the ZBCP is independent from a Zeeman field applied perpendicular to the N/S interface and disappear when the Zeeman field is parallel to the interface~\cite{Linder_2010}. The authors of Ref.~\cite{Linder_2010} note that the Majorana fermions spread along the interface separating the normal and the superconducting region.

By analysing the cooling power and  of a Josephson system based on 3DTI/$d$-wave junctions, it was found that the thermal conductance is profoundly sensitive to the components of the pairing state and that this can be varied by changing the length of the 3DTI region of the junction~\cite{HLi_2017,Salehi_2010}.

\section{Current experimental status}\label{Experiment}
\subsection{Experiments $s$-wave/graphene as the starting point}\label{exp_swave}
The first graphene-based Josephson junction was realized fourteen years ago~\cite{Heersche_2007} with an exfoliated graphene flake on top of SiO$_2$ and contacted by two Al electrodes. Since then, many groups realized and improved the quality of graphene Josephson junctions --- namely the contact transparency, and the graphene mobility. One crucial improvement is to use $h$-BN to encapsulate graphene which result in a higher graphene mobility~\cite{Dean}. This allowed to image the supercurrent distribution in the graphene layer close and far from the charge neutrality~\cite{Allen_2015}, revealing guided electronic mode along the edge close to charge neutrality. A second improvement is  to use superconducting electrodes with  high critical magnetic field such as Molybdenum Rhenium~\cite{Amet_2016} or Niobium Nitride~\cite{Lee2017}. This allowed to couple quantum Hall integer edge channels to superconducting leads, a promising step towards the realization of Majorana modes~\cite{Fu2008} or parafermions~\cite{Mong2014}, predicted to appear from the  coupling of  fractional quantum Hall edge channels to superconducting leads.
Whereas  high-quality contacts between graphene and $s$-wave superconductors  have been achieved, the techniques used  are  of little help to reach the same contact quality between graphene and a $d$-wave superconductors. As discussed earlier it is not possible to deposit a $d$-wave superconductor by evaporation on  a $h$-BN graphene stack. Indeed,  $d$-wave superconductor require specific growth conditions: specific substrate, high temperature, O$_2$ pressure, which are not compatible with the presence of graphene. The existing works combining graphene with $d$-wave superconductor~\cite{Di_Bernardo_2017,Perconte_2017,Perconte2020} have therefore used a wet transfer process to deposit a sheet of CVD grown graphene on top of a superconducting oxide. Although it achieves lower graphene mobility, the strength of this technique is its  scalability. An unexplored way, so far, would be to use an exfoliable $d$-wave superconductor to realize a $d$-wave superconductor graphene stack by the pick-up technique. This technique has been successfully applied to bilayer graphene with low-temperature superconductor~\cite{Efetov_2015} and high-temperature superconductor with topological insulator as detailed below.

\subsection{STM studies of $d$-wave superconductor/graphene proximity effect}\label{STM}
%
%
\begin{figure}[!t]
    \centering
    \includegraphics[width=0.95\textwidth]{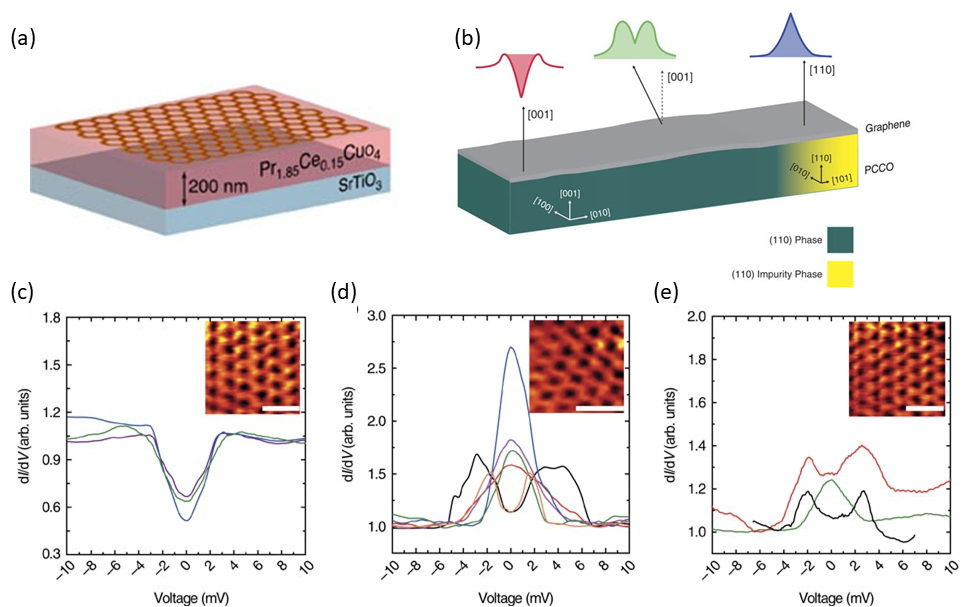}
    \caption{\label{fig:PCCO-Gr}STM studies of proximity induced superconductivity in graphene on top of a high-$T_\text{C}$ superconductor. (a) Sketch of the investigated samples, that consist of a single layer graphene sheet deposited on a 200~nm thick PCCO superconducting film that is grown on a STO substrate. The tunneling spectra is measured at different points by positioning the STM tip above the graphene layer. (b) Theoretical expectations for the tunneling conductance spectra in case of $p$-wave induced symmetry. The different projections of the p-wave symmetry result in characteristic spectral features that are correlated with the crystal orientation of the underling PCCO, as discussed in the main text. (c)-(e) show experimental tunneling conductance spectra which, depending on the tip position, can have a V-shape (a), a zero-bias conductance peak (d) or split ZBCPs. The inset show the topographic image of the areas where spectra are taken. Adapted from Ref.~\cite{Di_Bernardo_2017}.}
\end{figure}
%
%

The first reports of proximity induced superconductivity in graphene using cuprate superconductors were based on scanning tunneling microscopy (STM) experiments~\cite{Di_Bernardo_2017}. The  superconductor was Pr$_{2-x}$Ce$_x$CuO$_4$ (PCCO), whose critical temperature, $\sim20$~K is lower than the $T_\text{C}$ archetypal cuprate YBa$_2$Cu$_3$O$_7$ (YBCO) for which $T_\text{C}\sim 90$~K. The main characteristic of the studied electron-doped PCCO films, in addition to a relatively long coherence length $\sim$ 30 nm, is that the $d$-wave features are allegedly absent from the density of states because samples were in the dirty limit, that is, the mean free path is shorter that the coherence length. As shown in Fig.~\ref{fig:PCCO-Gr}(a), the proximity effects were studied in bilayers consisting of chemical vapor deposited (CVD) single-layer graphene (SLG) that is transferred onto 200 nm PCCO film using standard techniques~\cite{Bonaccorso2012}. STM was used to locally measure the differential conductance \emph{vs.} voltage bias V in areas where the surface topography made sure the presence of SLG as shown in Figs.~\ref{fig:PCCO-Gr}(c)-\ref{fig:PCCO-Gr}(e).  These experiments revealed gap-related spectral features that were only present when the critical temperature was well below the $T_\text{C}$ of the PCCO and gradually disappear as temperature was increased. Three main different types of behavior were observed, which are illustrated by the measurements displayed in Fig.~\ref{fig:PCCO-Gr}(c)-\ref{fig:PCCO-Gr}(e). The most common spectra (45 \% of them) contained V-shaped features, see Fig.~\ref{fig:PCCO-Gr}(c), that resemble those expected in bare PCCO. Other spectra presented very distinct features such as Zero-Bias Conductance Peaks (30 \% of the spectra as in Fig.~\ref{fig:PCCO-Gr}(d)) and split-ZBCPs (25 \% of the spectra as in Fig.~\ref{fig:PCCO-Gr}(e)) were observed that could not be found in bare PCCO. The authors ascribed the different spectral features to the emergence of $p$-wave superconductivity in the SLG due to proximity with PCCO. This conclusion is supported by  theoretical calculations of the density of states induced in graphene due to proximity with $d$-waves superconductors~\cite{Jiang2008,Linder_2009}. These theories suggest  that the different types of experimentally observed conductance curves can be explained if one considers that different projections of the $p$-paring were probed by the STM tip. This scenario was ascribed to the surface roughness of the PCCO film, which lead to local variations of the crystal orientation, each of them associated with a different $p$-wave projection in the SLG plane [see sketch in Fig.~\ref{fig:PCCO-Gr}(b)].

\subsection{Devices based on $d$-wave superconductors and graphene}\label{Device_Fabrication}

If STM experiments, as those discussed in the previous section, are a convenient approach to explore the proximity-induced superconducting correlations and density of states in graphene, they do not provide direct access to the conductance across the cuprate/graphene interface. In addition, the existing  STM experiments could not probe the length-scale over which the unconventional correlations penetrate into graphene~\cite{Di_Bernardo_2017}. Moreover, the geometry needed for STM  does not allow for an easy  gating of the graphene layer, and hence it  prevents for a  control of the  Fermi  level. Yet, as discussed in Section~\ref{Theory_dwave_graphene}, those aspects are crucial for understanding, tailoring and eventually exploiting high-$T_\text{C}$ superconducting proximity effects in graphene. A way of  tackling these issues is by fabrication of solid-state devices, similar to those used with  conventional low-$T_\text{C}$ superconductors (see Section~\ref{exp_swave}). Yet, as explained in Section~\ref{Dwave_material}, the growth of cuprates is  incompatible with some of the standard lithography techniques.
Because of the need to grow them on specific substrates, at very high deposition temperatures in an oxygen-rich atmosphere, cuprates cannot be grown on graphene.  Thus, the standard approach  consisting of the deposition of  low-$T_\text{C}$ superconducting electrodes on top of a graphene flake lying on a substrate, cannot be applied for the case of cuprates.  Instead, the cuprate film has to be epitaxially grown first on an appropriate substrate, \emph{e.g.} SrTiO$_3$ (STO), and then lithographic technique have been used 
to define the superconducting electrodes’ geometry, before a graphene sheet can be transferred onto them. Early attempts, by Sun \emph{et al.}~\cite{Sun2014a}, to implement such procedure via  conventional photolithography and dry etching techniques to define the YBCO electrodes on which graphene was subsequently transferred, demonstrated that the cuprate/graphene interfaces obtained in that way generally presented very low transparency, thus precluding proximity effects. The experiment suggested that exposure of YBCO to chemicals and resists degrades its superconducting and normal-state properties on the $c$-axis surface, leading to an insulating layer. The experiments by Sun \emph{et al.} nevertheless showed  hints of induced superconductivity in graphene in some of the measured devices. This was ascribed to a direct coupling between graphene and the YBCO’s CuO$_2$ planes at a local scale, which could be randomly achieved due to very rough surface of the used films.

%
%
\begin{figure}
    \centering
    \includegraphics[width=0.95\textwidth]{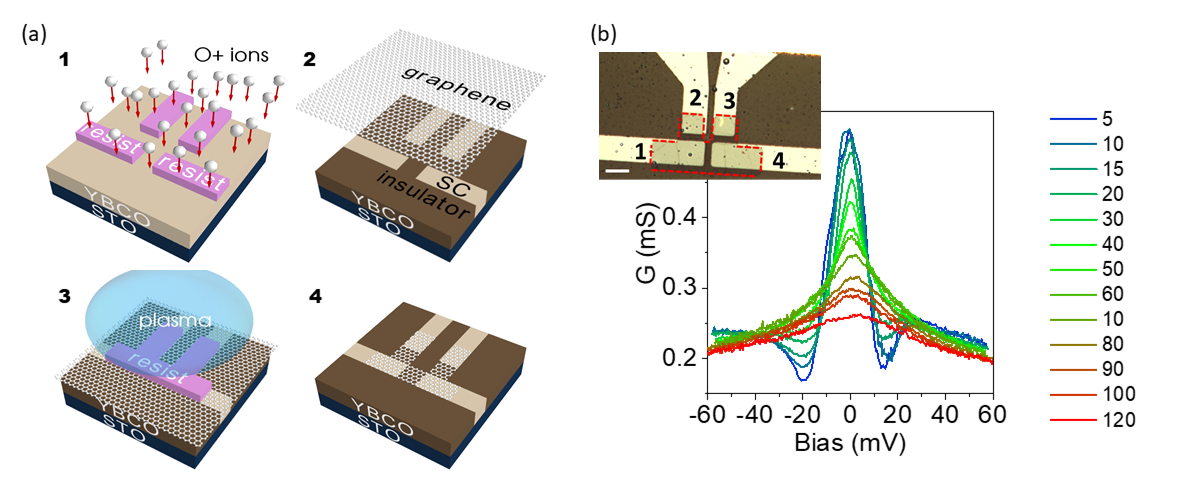}
    \caption{\label{fig:YBCO-Gr}Fabrication of solid-state devices for measuring the conductance across the $d$-wave/graphene interface. (a) The fabrication is based on 4 main steps. First, an Au (4~nm)/YBCO (50~nm) film is grown on SrTiO$_3$. In step 1, photolithography is used to define a resist mask through which the film is irradiated with 110 KeV O+ ions at fluence $5\times  10^{14}$~cm$^{-2}$. The irradiation makes the unprotected YBCO insulating, while the protected YBCO keeps superconducting (S) properties. In step 2,  Au is removed with a soft Ar$^+$ ion etching and the resist mask is desolved in acetone, after which a planar device is obtained in which four S electrodes (bright color) are separated by insulating YBCO (dark). In step 3, a single-layer graphene sheet is transferred onto the electrodes and subsequently patterned using photolithography and oxygen plasma to define a bridge between the S electrodes. Removal of the remaining resist yield the final device. (b) Shows a typical conductance spectrum in a device in which the YBCO/Au/graphene interface is highly transparent, which is characterized by a zero bias peak that grows as temperature is decreased below $T_\text{C}$. The inset shows a typical microscopy image on an actual device (scale bar is 10~$\mu$m). Adapted from Ref.~\cite{Perconte_2017}.}
\end{figure}
%
%
An alternative fabrication method to improve the coupling between graphene and $d$-wave superconductivity was developed by  Perconte~\emph{et al.}~\cite{Perconte_2017} 
and it is schematically shown in Figure~\ref{fig:YBCO-Gr} (a).  The first feature of the new approach is that YBCO films are covered \emph{in situ} (that is, right after its growth in a pulsed laser deposition system and without breaking the vacuum of the deposition chamber) by a few nm thick Au capping layer. This layer protects the YBCO surface during the subsequent fabrication steps, preserving its superconducting properties. Importantly the Au/cuprate transparency is very high as demonstrated in Ref.~\cite{Chang2004} for Au/BSCCO interface and ~\cite{Perconte_2017} for Au/YBCO interface. Moreover, because the capping layer of Au is one order of magnitude thinner than the mean free path in this material, one expects $d$-wave superconducting correlations to propagate essentially undisturbed across it~\cite{Tsuneto_1962,Balatsky_2006}. A second feature of this fabrication method  is that, in order to define the device geometry,  in particularly the YBCO electrodes, masked ion irradiation is used instead of etching.  Irradiation turns the exposed YBCO amorphous, and thus insulating, while the masked YBCO areas keep superconducting properties. After removal of the thin Au layer from the areas exposed to the ion bombardment (using a mild Ar$^+$ plasma) and of the resist mask, a pattern of superconducting YBCO/Au electrodes embedded in an insulating YBCO matrix is obtained. The device obtained following this procedure is therefore nearly planar (only a ~ 4 nm topography step due to the Au thickness) which contrasts with the case of etched electrodes that yield large steps comparable with the YCBO thickness. Thus, in the irradiated devices, graphene lies on a flatter surface. The rest of the fabrication process is standard and includes, after the transfer of CVD graphene, a lithography and O$_2$ plasma etching to shape graphene into a bridge connecting the YBCO/Au electrodes. In the  device shown in Fig.~\ref{fig:YBCO-Gr}, the graphene bridge is contacted with a total of four electrodes. This  allowed  two, three or four probe measurements. Electrostatic doping of the graphene can be done either by using the high-$k$ dielectric STO substrate as a backgate, or an AlO$_x$ top-gate that is deposited after completing the lithography steps 1-4, as shown in Fig.~\ref{fig:YBCO-Gr}(a). Figure~\ref{fig:YBCO-Gr}(b) shows a three-probe conductance measurement obtained in this type of device. One observes that, as temperature is decreased below  $T_\text{C}$, the low-bias conductance is gradually enhanced to form a peak that extends over an energy scale comparable to the expected superconducting gap of YBCO. The authors of~\cite{Perconte_2017} explained this behavior by assuming a highly transparent  Au/graphene interface, which allows superconductivity to be proximity-induced in the graphene that lies directly on the superconducting YBCO/Au contact. In this scenario, as  theoretically explored~\cite{Linder_2008}, the measured conductance essentially corresponds to the conductance across the interface that separates superconducting graphene (that lying directly on the YBCO/Au electrodes) and normal graphene (the one lying on insulating YBCO).

%
%
\begin{figure}
    \centering
    \includegraphics[width=0.95\textwidth]{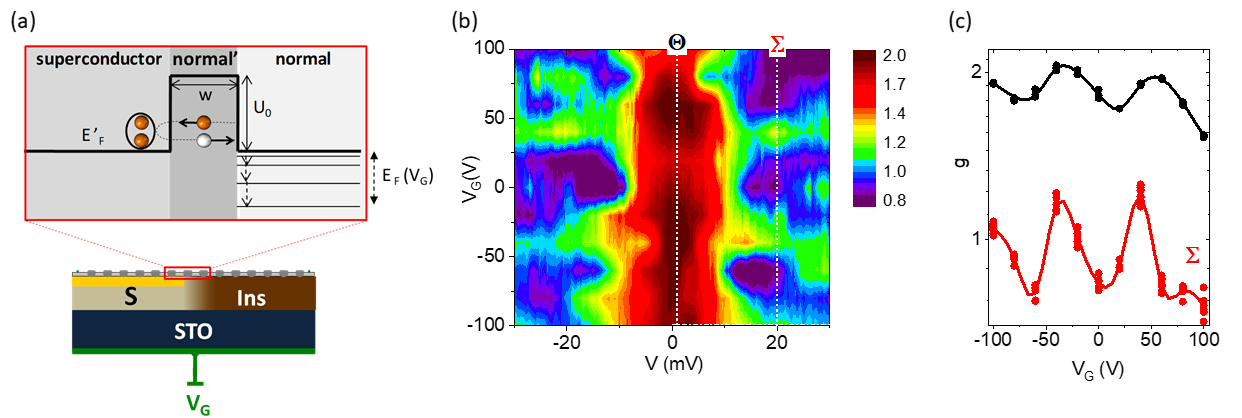}
    \caption{\label{fig:Klein-tunneling}Superconducting Klein-like tunneling. (a) Schematic of the model used to describe the conductance across the interface between proximitized graphene (lying on superconducting YBCO/Au) and graphene lying on insulating YBCO, which is back gated across the dielectric STO substrate as shown in the lower cartoon. A potential step of width w and height $U_0$ defines a region of normal graphene (N$^{\prime}$) that separates superconducting graphene with Fermi energy $E_\text{F}^{\prime}$ from normal graphene (N) with gate-tunable Fermi energy $E_\text{F}$($V _\text{G}$). (b) Contour plot of the normalized conductance (colour scale) measured as a function of the bias $V$ across the junction and gate voltage $V_\text{G}$. (c) Conductance as a function of $V_\text{G}$ at fixed bias $V$. These profiles correspond to the ‘cuts’ of the plot in (b), and show the oscillatory behavior of the conductance as a function of the gate voltage. Adapted from Ref.~\cite{Perconte_2017}.}
\end{figure}
%
%

The above assumption was confirmed via backgating experiments~\cite{Perconte_2017},  with the geometry shown in Fig.~\ref{fig:Klein-tunneling}(a).  The electrostatic doping only occurs in the portion of graphene lying on insulating YBCO because the electric field is screened elsewhere by the (super)conducting YBCO/Au. In the reported measurements, upon application of a gate voltage, the spectra is strongly modified as compared to the case shown in Fig.~\ref{fig:YBCO-Gr}(b). This can be seen in Fig.~\ref{fig:Klein-tunneling}(b), which shows a contour plot  of the conductance (color code) \emph{vs.} bias ($x$-axis) as a function of the applied gate voltage ($y$-axis). In particular, the relative height of low-bias conductance enhancement (red area in the color plot), as well as the background conductance level, is periodically modulated by the gate voltage, as shown in Fig.~\ref{fig:Klein-tunneling}(c).  
%
%
\begin{figure}[!t]
    \centering
    \includegraphics[width=0.95\textwidth]{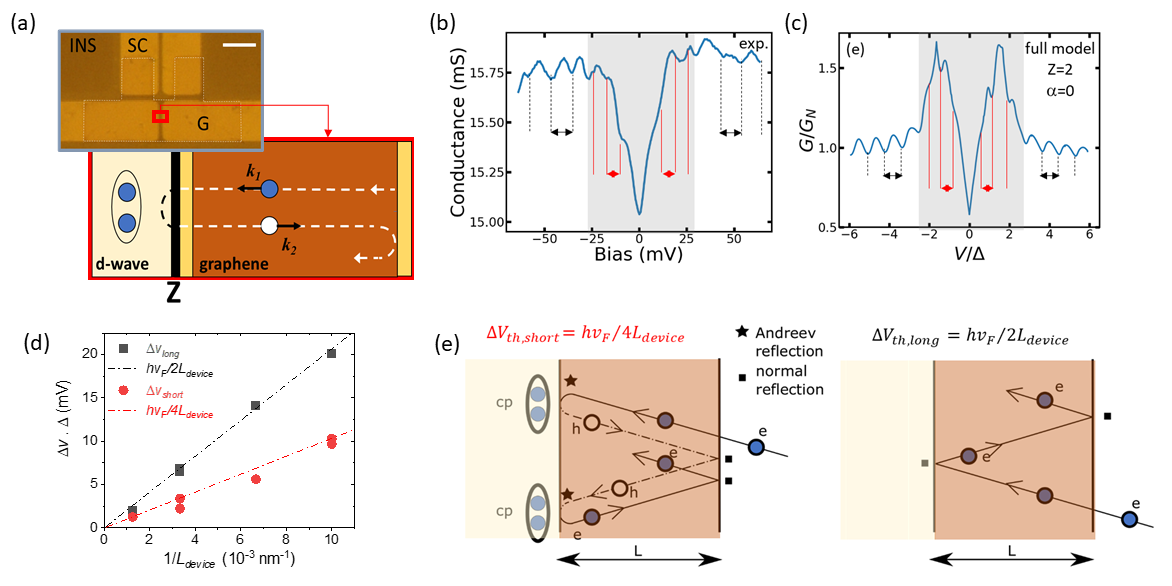}
    \caption{\label{fig:long-range}Long-range propagation and interference of $d$-wave Andreev pairs. (a) Microscope image of a graphene cavity formed in between two superconducting electrodes (SC) bridged by an overlaying single-layer graphene sheet  on an insulating substrate. The scale bar is 10~$\mu$m.  The scheme is a cartoon of the cavity. The graphene doping and thus its Fermi energy depends on whether it lies on insulating (brown) or superconducting (yellow) YBCO, which creates the cavity in which the electron and Andreev pair interferences occur. The YBCO/Au/graphene interface measured in series with the cavity is characterized by a barrier strength $Z$. (b) Typical experimental conductance vs. bias across a cavity, which shows oscillations with two different periods as shown by the black (long period) and red (short period) arrows. (c) Numerical simulation of the cavity’s conductance using the model described in the main text. (d) Oscillations period extracted from the modelling as a function of the inverse of the cavity’s length. The ratio between the long (black) and short (red) oscillations period is 2. (e) Scheme of the different types of interferences responsible for the conductance oscillations. Interferences in a cavity in proximity with a superconductor. In the left case, an electron is Andreev-reflected at the SC/cavity interface as a hole propagates back to the other end. There, it is normal-reflected towards the SC/cavity interface, where it undergoes again Andreev-reflection. This allows the interference that yield the short period oscillations. In the right case, the electron is normal-reflected at the SC/cavity, travels back to the other cavity interface where it is also normal-reflected- which produces the long period resonances. The predominance of one or the other type of resonances depends on the ratio between Andreev reflection and normal reflection at the SC interface and thus, for a given interface transparency, on the energy of the involved electrons. Adapted from Ref.~\cite{Perconte2020}.}
\end{figure}
%
%
This behavior can be explained considering doping profile across the interface between superconducting and non superconducting graphene, as sketched in the top part of Fig.~\ref{fig:Klein-tunneling}(a). The Fermi level in the superconducting graphene is pinned because here gate effect are screened by the YBCO/Au. Contrarily, in normal graphene (on top of insulating YBCO) the Fermi level varies upon gating, and consequently the Fermi vector $k_\text{F}$ of the carriers flowing across the interface does too. The authors of Ref.~\cite{Perconte_2017} assumed  a step-like profile  of the Fermi energy, as shown in Fig.~\ref{fig:Klein-tunneling}(a) due to edge effects (\emph{e.g.} fringe fields associated to the backgate geometry, gradient of electronic properties between irradiated and non/irradiated YBCO\ldots). Such a profile results in a potential energy step $U_0$ for electrons between superconducting and normal graphene. This situation is analogous to that in graphene $n–p–n$ or $p–n–p$ junctions, in which  Klein tunneling is observed~\cite{Katsnelson_2006,Beenakker_2008,Huard2007,Young_2009}. But  now, because of the presence of superconducting correlations,  transport at low-bias involves Andreev pairs. In this scenario, interference between Andreev particles (or electrons for bias above the energy-gap) are expected in the central (normal) region~\cite{Linder_2008}, whose  width $w$ was estimated at $w\sim60$~nm, as particles undergo reflections and travel back and forth between its two ends. This leads to a modulation of the conductance that  depends on the phase picked by the charge carriers when travelling across that central region, which is $ \chi = -w k_\text{F}$. Thus, upon gating, the induced change of $k_\text{F}$ leads to a periodic modulation of the conductance, which explain the experiments in Fig.~\ref{fig:Klein-tunneling}(b) and~\ref{fig:Klein-tunneling}(c).  All these findings were supported by a numerical calculation of the conductance using the model of Ref.~\cite{Linder_2008}. A good agreement between experiment and theory was found~\cite{Perconte_2017}.

The above experiments demonstrated the superconducting proximity effect across the $d$-wave/graphene interface and its modulation via  gate voltage.  However, from those experiments one could not determine the (lateral) length scale over which the induced superconducting correlations propagate into graphene.  To investigate this issue, similar devices were fabricated but with a much shorter distance between the superconducting electrodes~\cite{Perconte2020}. Such device is shown  Fig.~\ref{fig:long-range}(a). The distance between the superconducting electrodes  is of the order of a few hundred nm. If this distance is  shorter than the coherence length for Andreev pairs and mean-free path of the electrons, then both Andreev pairs and normal electrons can coherently propagate and suffer reflections at both ends of the normal graphene. In such a case the device behaves as a “cavity” in which reflections at its ends leads to Andreev pair and electron confinement, and ultimately to interference effects.  Such interferences, illustrated in the schemes of Fig.~\ref{fig:long-range}(e) are respectively analogous to the De Gennes-Saint James~\cite{de_Gennes_1963} or McMillan-Rowell oscillations~\cite{Rowell_1966} for Andreev pairs, and to Fabry-P\'erot interferences for electrons~\cite{Liang_2001,Miao_2007,Young_2009,Campos_2012,Allen_2017}. 
The existence of these interference effects was confirmed in the experiment of Ref.~\cite{Perconte2020}.   They manifest in the  conductance measurements via two distinct families of oscillations as a function of the bias voltage across the cavity [Fig.~\ref{fig:long-range}(b)]. The shorter period oscillations correspond to Andreev pair interference effects, and are dominant at low bias. The oscillations stemming from electron interference present a doubling of the period and are dominant at bias above the superconducting energy gap.  To support this interpretation of the experiments, numerical simulations of the conductance were performed based on Blonder-Tinkham-Klapwijk  model generalized to $d$-wave superconductors~\cite{Kashiwaya_1996,Wei_1998} (see Section~\ref{Theory}) to deal with YBCO/Au/graphene interfaces of finite transparency and on the proximitized graphene homojunction~\cite{Linder_2007,Linder_2008} to model the cavity’s conductance in series. This approach allowed reproducing numerically the main features of the experimental curves, including not only the oscillations,  but also the background conductance shape, which is determined by the transparency of the YBCO/Au/graphene interface (described by the barrier strength parameter $Z$ of the BTK theory). The experiments were carried in devices with different cavity’s lengths, up to 800~nm, which allowed verifying the consistency of the analysis as the oscillations periods extracted from the analysis of the experimental curves scaled with the inverse of the cavity’s length as expected from the theory, see Fig.~\ref{fig:long-range}(d). In summary, these experiments demonstrated the long-range (hundreds of nm) of the superconducting correlations induce in graphene by proximity with a $d$-wave superconductor. The model used considered that the induced correlations retain the original $d$-wave symmetry of the cuprate, and no indication of a different symmetry could be found from the analysis. 

\subsection{Experiments on topological insulators and d-wave superconductors}\label{exp_TI}
%
%
\begin{figure}[!t]
    \centering
    \includegraphics[width=0.95\textwidth]{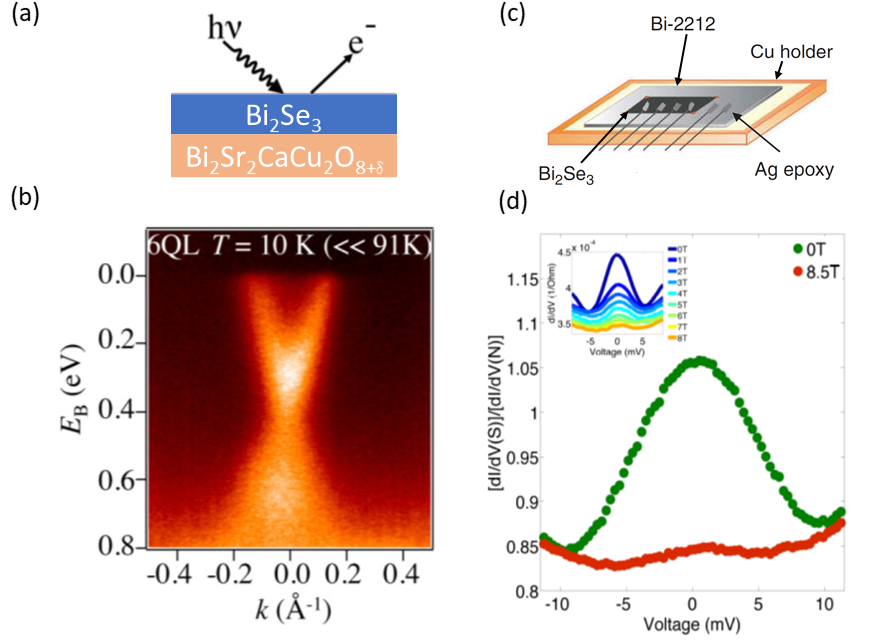}
    \caption{\label{fig_Ti_supra} Examples of hybrid structures between 3D topological insulators and $d$-wave superconductors. (a) Bi$_2$Se$_3$ is grown on top of BSCCO by molecular beam epitaxy, (b) the corresponding ARPES spectra shows a Dirac like energy dispersion, adapted from Ref.~\cite{Xu2014}. (c) Conversely BSCCO islands have been grown on top of Bi$_2$Te$_2$Se (d) the heterostructure conductance show superconductivity related features indicating proximity effect, adapted from Ref.~\cite{Zareapour2012}.}
\end{figure}
%
%
The quest to evidence and manipulate Majorana fermions is a very hot topic~\cite{Qi2011,Beenakker2015} motivating most of the study of superconductor/topological insulator devices --- although other prompt applications of these devices exists, for example for enhanced photoluminescence~\cite{Asano2009, Hayat2012,Hayat2014}. Like for graphene, experiments between topological insulator and $s$-wave superconductor have been focused on the transport properties of Josephson junctions~\cite{Culcer_2020}. These include HgTe/CdTe in contact with Al leads which showed missing Shapiro steps interpreted as induced superconductivity in the quantum spin Hall edges~\cite{Bocquillon2017}. Both 2D and 3D topological insulators were studied in this work, and the differences between both cases was discussed extensively in Ref.~\cite{Bocquillon2018}. The three-dimensional topological insulator Bi$_2$Se$_3$ has been contacted with Al leads~\cite{Sacepe2011} showing gate tunable supercurrent as well as Nb superconducting leads showing multiple Andreev reflections~\cite{Kurter2014}. More recently, Josephson junction with Al superconducting electrodes and a Bi$_2$Se$_3$ spacer showed a critical current minimum at zero magnetic field, possibly indicating a $p$-wave induced superconductivity~\cite{kunakova2020topological, charpentier2017induced,galletti2014influence}. In a recent paper on arxiv~\cite{rosen2021fractional}, the group of David Goldhaber-Gordon reported a new technique yielding transparent interface between the 3D topological insulator (Bi$_{0.4}$Sb$_{0.6}$)$_2$Te$_3$ and the superconducting alloy PdTe, as the authors note, their results are compatible with Majorana fermions but are not an evidence for their existence. Dirac semimetals have also been used as spacers in $s$-wave Josephson junctions, Cd$_3$As$_2$~\cite{Wang2018,Yu2020}, Bi$_{1-x}$Sb$_x$~\cite{li20184pi,li2019zeeman}, with evidence for topological superconductivity such as $4 \pi$ periodic supercurrent or a Zeeman field transition between a normal and a $\pi$ Josephson junction. 

he work on $d$-wave superconductors combined with topological insulators is so far scarce. Theoretically,  it has been proposed that the elementary excitations at the interface between a $d$-wave superconductor and a topological insulator are Majorana fermions~\cite{Linder_2010}. The few experiments combining topological insulators with $d$-wave superconductors are summarized in Fig.~\ref{fig_Ti_supra}. The two techniques exploited are  point contact spectroscopy measurements~\cite{Zareapour2014} and Angle-Resolved Photoemission Spectroscopy (ARPES)~\cite{Xu2014}. These two approaches led to contradicting results~\cite{Yilmaz2014,Xu2014,Zareapour2012} concerning the possibility to induce  $d$-wave superconductivity  in topological insulators. Using topological insulators grown on  a $d$-wave superconductor and ARPES  the groups of T. Valla and Z. Hasan~\cite{Yilmaz2014,Xu2014} concluded that the proximity effect is prohibited in these structures due to Fermi surface mismatch.  Burch's group, based on exfoliated d-wave/TI devices, transport experiments~\cite{Zareapour2016} and theory~\cite{Zareapour2017} showed the opposite.  To our knowledge, no other topological insulator  in proximity with $d$-wave superconductors has been studied so fat, which might be due to the growth conditions making it difficult to combine both class of materials. 
Among the different topological insulators WTe$_2$~\cite{Fei2017} seems a very suitable candidate to be brought into contact with a $d$-wave superconductor, because it  can be exfoliated and transferred onto different materials~\cite{Ma2019}. Furthermore, monolayer WTe$_2$ has recently been shown to display superconductivity~\cite{Fatem} and the pairing symmetry is so far unknown. One can only speculate that in the case of a $d$-wave pairing symmetry, one would be able to realize WTe$_2$ (SC)/WTe$_2$ junction by local gating~\cite{Sajadi2018}. On a final note, it is worth pointing out that in, general, the smoking gun signatures of superconductivity mediated by topological edge states channels did not reach a consensus yet~\cite{DeVries2018a},  and that these difficulties will of course apply to the case of unconventional superconductors.

\section{Summary and Outlook}\label{Conclusion}
The combination of $d$-wave cuprate superconductors with Dirac materials bears great fundamental interest and potential for technological applications. As we discussed in the theory section, the isotropic pairing results in a large density of zero-energy superconducting quasiparticles and shows a non-uniform superconducting phase. This results in directional effects,  dictated by the orientation between the cuprate's and the Dirac material's crystalline axes within the $x$-$y$ plane. Unique features are expected, such as the generation of unconventional pairing symmetries, the large dominance of low-energy Andreev bound states in the conductance across the cuprate/Dirac material interface, or the tantalizing possibility of realizing  $\pi$ Josephson junctions with present gate tunability, potentially at relatively high temperatures of the order of tens of K. So far, very few experimental works have been conducted, to a great extent because the realisation of junctions between cuprates and Dirac materials which are sufficiently transparent to allow for the proximity effect is far more challenging than with conventional, low-$T_\text{C}$ superconductors. Despite this, graphene as a prototypical Dirac material has been successfully combined with different cuprates. Spectroscopic measurements of graphene on PCCO using scanning tunneling microscopy found signatures of proximity-induced unconventional $p$-wave pairing in graphene~\cite{Di_Bernardo_2017}. 
The development of \textit{ad hoc} nanolithography approaches enabled the fabrication of YBCO/graphene devices with transparent interfaces~\cite{Perconte_2017} in which the tunneling of Andreev pairs can be modulated by an electrostatic gate, analogously as in the Klein tunneling of normal electrons. The same type of devices have allowed demonstrating that  $d$-wave superconducting correlation can coherently propagate into graphene over distances of the order of hundreds of~nm~\cite{Perconte2020}. These studies are promising steps towards the realization of $d$-wave Josephson junction with a Dirac material spacer~\cite{Linder_2008}, which in practical terms would require further enhancing and reliably controlling the interfaces' transparency. Further prospects include the demonstration of high-temperature Josephson effects across topological insulators and other materials from the ever-growing library of exfoliated 2D materials~\cite{Liu2016,Novoselov2016}. As described above, a few promising results already exist based on experiments with bilayer heterostructures that combine cuprates with topological insulators ~\cite{Yilmaz2014,Xu2014,Zareapour2012}, and novel effects have been found in junctions based on the exfoliated cuprate bismuth strontium calcium copper oxide (BSCCO)~\cite{Zhu2021,Zhao2021}, which opens worthwhile avenue towards BSCCO–graphene and BSCCO-3DTI devices.

\medskip

\medskip
\textbf{Acknowledgements} \par 
Work at Unit\'e Mixte de Physique CNRS/Thales supported by ERC grant N. 647100 "SUSPINTRONICS", French ANR grant ANR-17-CE30-0018-04 "OPTOFLUXONICS", COST action "Nanocohybri", European Union's H2020 Excellent Science Grant  N. 881603 "GRAPHENE CORE 3", Labex NanoSaclay ANR-10-LABX-0035 and JSPS core-to-core program "A. Advanced Research Networks". The work of S.B. and D.B. is support from Ministerio de Ciencia e Innovaci\'on  (MICINN) through Project N. PID2020-114252GB-I00 (SPIRIT) and  PID2020-120614GB-I00 (ENACT), and by the Transnational Common Laboratory $Quantum-ChemPhys$. Further, D.B. and S.B. ackowledge the funding from the Basque Government's IKUR initiative on Quantum technologies (Department of Education). F.S.B thanks Prof. Bj\"orn Trauzettel for his hospitality at W\"urzburg University,  the A. v. Humboldt Foundation for financial support, and EU's the Horizon 2020 Research and
Innovation Program fro funding under Grant Agreement No. 800923 (SUPERTED).

\medskip

%
\bibliographystyle{MSP}
\bibliography{bibliography}

\begin{thebibliography}{100}
\providecommand{\url}[1]{\texttt{#1}}
\providecommand{\urlprefix}{URL }

\bibitem{Castro_Neto_2009}
A.~H.~C. Neto, F.~Guinea, N.~M.~R. Peres, K.~S. Novoselov, A.~K. Geim,
\newblock \emph{Rev. Mod. Phys.} \textbf{2009}, \emph{81}, 1 109.

\bibitem{Beenakker_2008}
C.~W.~J. Beenakker,
\newblock \emph{Rev. Mod. Phys.} \textbf{2008}, \emph{80} 1337.

\bibitem{Heersche_2007}
H.~B. Heersche, P.~Jarillo-Herrero, J.~B. Oostinga, L.~M.~K. Vandersypen, A.~F.
  Morpurgo,
\newblock \emph{Nature} \textbf{2007}, \emph{446}, 7131 56.

\bibitem{Allen_2015}
M.~T. Allen, O.~Shtanko, I.~C. Fulga, A.~R. Akhmerov, K.~Watanabe,
  T.~Taniguchi, P.~Jarillo-Herrero, L.~S. Levitov, A.~Yacoby,
\newblock \emph{Nat. Phys.} \textbf{2015}, \emph{12}, 2 128.

\bibitem{Ben_Shalom_2015}
M.~B. Shalom, M.~J. Zhu, V.~I. Fal'ko, A.~Mishchenko, A.~V. Kretinin, K.~S.
  Novoselov, C.~R. Woods, K.~Watanabe, T.~Taniguchi, A.~K. Geim, J.~R. Prance,
\newblock \emph{Nat. Phys.} \textbf{2015}, \emph{12}, 4 318.

\bibitem{Amet_2016}
F.~Amet, C.~T. Ke, I.~V. Borzenets, J.~Wang, K.~Watanabe, T.~Taniguchi, R.~S.
  Deacon, M.~Yamamoto, Y.~Bomze, S.~Tarucha, G.~Finkelstein,
\newblock \emph{Science} \textbf{2016}, \emph{352}, 6288 966.

\bibitem{Ortlepp2006}
T.~Ortlepp, Ariando, O.~Mielke, C.~J. Verwijs, K.~F. Foo, H.~Rogalla, F.~H.
  Uhlmann, H.~Hilgenkamp,
\newblock \emph{Science} \textbf{2006}, \emph{312}, 5779 1495.

\bibitem{Koelle1999a}
D.~Koelle, R.~Kleiner, F.~Ludwig, E.~Dantsker, J.~Clarke,
\newblock \emph{Rev. Mod. Phys.} \textbf{1999}, \emph{71} 631.

\bibitem{Ouanani2016}
S.~Ouanani, J.~Kermorvant, C.~Ulysse, M.~Malnou, Y.~Lemaître, B.~Marcilhac,
  C.~Feuillet-Palma, N.~Bergeal, D.~Crété, J.~Lesueur,
\newblock \emph{Supercond. Sci. Technol.} \textbf{2016}, \emph{29} 094002.

\bibitem{Swiecicki2012}
I.~Swiecicki, C.~Ulysse, T.~Wolf, R.~Bernard, N.~Bergeal, J.~Briatico,
  G.~Faini, J.~Lesueur, J.~E. Villegas,
\newblock \emph{Phys. Rev. B} \textbf{2012}, \emph{85} 224502.

\bibitem{Di_Bernardo_2017}
A.~D. Bernardo, O.~Millo, M.~Barbone, H.~Alpern, Y.~Kalcheim, U.~Sassi, A.~K.
  Ott, D.~D. Fazio, D.~Yoon, M.~Amado, A.~C. Ferrari, J.~Linder, J.~W.~A.
  Robinson,
\newblock \emph{Nat. Comm.} \textbf{2017}, \emph{8}, 1 14024.

\bibitem{Perconte_2017}
D.~Perconte, F.~A. Cuellar, C.~Moreau-Luchaire, M.~Piquemal-Banci, R.~Galceran,
  P.~R. Kidambi, M.-B. Martin, S.~Hofmann, R.~Bernard, B.~Dlubak, P.~Seneor,
  J.~E. Villegas,
\newblock \emph{Nat. Phys.} \textbf{2017}, \emph{14}, 1 25.

\bibitem{Perconte2020}
D.~Perconte, K.~Seurre, V.~Humbert, C.~Ulysse, A.~Sander, J.~Trastoy, V.~Zatko,
  F.~Godel, P.~R. Kidambi, S.~Hofmann, X.~P. Zhang, D.~Bercioux, F.~S.
  Bergeret, B.~Dlubak, P.~Seneor, J.~E. Villegas,
\newblock \emph{Phys. Rev. Lett.} \textbf{2020}, \emph{125}, 8 087002.

\bibitem{Zareapour2017}
P.~Zareapour, A.~Hayat, S.~Y.~F. Zhao, M.~Kreshchuk, Z.~Xu, T.~S. Liu, G.~D.
  Gu, S.~Jia, R.~J. Cava, H.-Y. Yang, Y.~Ran, K.~S. Burch,
\newblock \emph{New J. Phys.} \textbf{2017}, \emph{19}, 4 043026.

\bibitem{Novoselov2005}
K.~S. Novoselov, D.~Jiang, F.~Schedin, T.~J. Booth, V.~V. Khotkevich, S.~V.
  Morozov, A.~K. Geim,
\newblock \emph{Proc. Natl. Acad. Sci. U.S.A.} \textbf{2005}, \emph{102}, 30
  10451.

\bibitem{Ferrari2015}
A.~C. Ferrari, F.~Bonaccorso, V.~Fal'ko, K.~S. Novoselov, S.~Roche,
  P.~Bøggild, S.~Borini, F.~H. Koppens, V.~Palermo, N.~Pugno, J.~A. Garrido,
  R.~Sordan, A.~Bianco, L.~Ballerini, M.~Prato, E.~Lidorikis, J.~Kivioja,
  C.~Marinelli, T.~Ryhänen, A.~Morpurgo, J.~N. Coleman, V.~Nicolosi,
  L.~Colombo, A.~Fert, M.~Garcia-Hernandez, A.~Bachtold, G.~F. Schneider,
  F.~Guinea, C.~Dekker, M.~Barbone, Z.~Sun, C.~Galiotis, A.~N. Grigorenko,
  G.~Konstantatos, A.~Kis, M.~Katsnelson, L.~Vandersypen, A.~Loiseau,
  V.~Morandi, D.~Neumaier, E.~Treossi, V.~Pellegrini, M.~Polini, A.~Tredicucci,
  G.~M. Williams, B.~H. Hong, J.~H. Ahn, J.~M. Kim, H.~Zirath, B.~J.~V. Wees,
  H.~V.~D. Zant, L.~Occhipinti, A.~D. Matteo, I.~A. Kinloch, T.~Seyller,
  E.~Quesnel, X.~Feng, K.~Teo, N.~Rupesinghe, P.~Hakonen, S.~R. Neil,
  Q.~Tannock, T.~Löfwander, J.~Kinaret,
\newblock \emph{Nanoscale} \textbf{2015}, \emph{7}, 11 4598.

\bibitem{Gong2012}
L.~Gong, R.~J. Young, I.~A. Kinloch, I.~Riaz, R.~Jalil, K.~S. Novoselov,
\newblock \emph{{ACS} Nano} \textbf{2012}, \emph{6}, 3 2086.

\bibitem{Lee2008}
C.~Lee, X.~Wei, J.~W. Kysar, J.~Hone,
\newblock \emph{Science} \textbf{2008}, \emph{321}, 5887 385.

\bibitem{Balandin2008}
A.~A. Balandin, S.~Ghosh, W.~Bao, I.~Calizo, D.~Teweldebrhan, F.~Miao, C.~N.
  Lau,
\newblock \emph{Nano Lett.} \textbf{2008}, \emph{8}, 3 902.

\bibitem{Zhu2010}
Y.~Zhu, S.~Murali, W.~Cai, X.~Li, J.~W. Suk, J.~R. Potts, R.~S. Ruoff,
\newblock \emph{Adv. Mater.} \textbf{2010}, \emph{22}, 35 3906.

\bibitem{Nair2008}
R.~R. Nair, P.~Blake, A.~N. Grigorenko, K.~S. Novoselov, T.~J. Booth,
  T.~Stauber, N.~M.~R. Peres, A.~K. Geim,
\newblock \emph{Science} \textbf{2008}, \emph{320}, 5881 1308.

\bibitem{Bunch2008}
J.~S. Bunch, S.~S. Verbridge, J.~S. Alden, A.~M. van~der Zande, J.~M. Parpia,
  H.~G. Craighead, P.~L. McEuen,
\newblock \emph{Nano Lett.} \textbf{2008}, \emph{8}, 8 2458.

\bibitem{Goerbig_2011}
M.~O. Goerbig,
\newblock \emph{Rev. Mod. Phys.} \textbf{2011}, \emph{83}, 4 1193.

\bibitem{Wallace_1947}
P.~R. Wallace,
\newblock \emph{Phys. Rev.} \textbf{1947}, \emph{71}, 9 622.

\bibitem{Torres_2013}
L.~E.~F. Torres, J.~C. Charlier, S.~Roche,
\newblock \emph{Introduction to graphene-based nanomaterials: from electronic
  structure to quantum transport},
\newblock Cambridge university press, \textbf{2013}.

\bibitem{Novoselov2004}
K.~S. Novoselov, A.~K. Geim, S.~V. Morozov, D.~Jiang, Y.~Zhang, S.~V. Dubonos,
  I.~V. Grigorieva, A.~A. Firsov,
\newblock \emph{Science} \textbf{2004}, \emph{306}, 5696 666.

\bibitem{Chen2008}
J.-H. Chen, C.~Jang, S.~Xiao, M.~Ishigami, M.~S. Fuhrer,
\newblock \emph{Nat. Nanotechnol.} \textbf{2008}, \emph{3}, 4 206.

\bibitem{Morozov2008}
S.~V. Morozov, K.~S. Novoselov, M.~I. Katsnelson, F.~Schedin, D.~C. Elias,
  J.~A. Jaszczak, A.~K. Geim,
\newblock \emph{Phys. Rev. Lett.} \textbf{2008}, \emph{100}, 1.

\bibitem{Du2008}
X.~Du, I.~Skachko, A.~Barker, E.~Y. Andrei,
\newblock \emph{Nat. Nanotechnol.} \textbf{2008}, \emph{3}, 8 491.

\bibitem{Neugebauer2009}
P.~Neugebauer, M.~Orlita, C.~Faugeras, A.-L. Barra, M.~Potemski,
\newblock \emph{Phys. Rev. Lett.} \textbf{2009}, \emph{103}, 13 136403.

\bibitem{Mohiuddin2008}
T.~M.~G. Mohiuddin, E.~Hill, D.~Elias, A.~Zhukov, K.~Novoselov, A.~Geim,
\newblock \emph{IEEE Trans. Magn.} \textbf{2008}, \emph{44}, 11 2624.

\bibitem{Blake2007}
P.~Blake, E.~W. Hill, A.~H.~C. Neto, K.~S. Novoselov, D.~Jiang, R.~Yang, T.~J.
  Booth, A.~K. Geim,
\newblock \emph{Appl. Phys. Lett.} \textbf{2007}, \emph{91}, 6 063124.

\bibitem{Berger2004}
C.~Berger, Z.~Song, T.~Li, X.~Li, A.~Y. Ogbazghi, R.~Feng, Z.~Dai, A.~N.
  Marchenkov, E.~H. Conrad, P.~N. First, W.~A. de~Heer,
\newblock \emph{J. Phys. Chem. B} \textbf{2004}, \emph{108}, 52 19912.

\bibitem{Obraztsov2009}
A.~N. Obraztsov,
\newblock \emph{Nat. Nanotechnol.} \textbf{2009}, \emph{4}, 4 212.

\bibitem{Chen2011}
S.~Chen, W.~Cai, R.~D. Piner, J.~W. Suk, Y.~Wu, Y.~Ren, J.~Kang, R.~S. Ruoff,
\newblock \emph{Nano Lett.} \textbf{2011}, \emph{11}, 9 3519.

\bibitem{Bae2010}
S.~Bae, H.~Kim, Y.~Lee, X.~Xu, J.-S. Park, Y.~Zheng, J.~Balakrishnan, T.~Lei,
  H.~R. Kim, Y.~I. Song, Y.-J. Kim, K.~S. Kim, B.~Ozyilmaz, J.-H. Ahn, B.~H.
  Hong, S.~Iijima,
\newblock \emph{Nat. Nanotechnol.} \textbf{2010}, \emph{5}, 8 574.

\bibitem{Kidambi_2012}
P.~R. Kidambi, C.~Ducati, B.~Dlubak, D.~Gardiner, R.~S. Weatherup, M.-B.
  Martin, P.~Seneor, H.~Coles, S.~Hofmann,
\newblock \emph{J. Phys. Chem. C} \textbf{2012}, \emph{116}, 42 22492.

\bibitem{Weatherup2012}
R.~S. Weatherup, B.~Dlubak, S.~Hofmann,
\newblock \emph{ACS Nano} \textbf{2012}, \emph{6}, 11 9996.

\bibitem{Mzali_2016}
S.~Mzali, A.~Montanaro, S.~Xavier, B.~Servet, J.-P. Mazellier, O.~Bezencenet,
  P.~Legagneux, M.~Piquemal-Banci, R.~Galceran, B.~Dlubak, P.~Seneor, M.-B.
  Martin, S.~Hofmann, J.~Robertson, C.-S. Cojocaru, A.~Centeno, A.~Zurutuza,
\newblock \emph{Appl. Phys. Lett.} \textbf{2016}, \emph{109}, 25 253110.

\bibitem{Kang2012}
J.~Kang, D.~Shin, S.~Bae, B.~H. Hong,
\newblock \emph{Nanoscale} \textbf{2012}, \emph{4}, 18 5527.

\bibitem{Wehling_2014}
T.~Wehling, A.~Black-Schaffer, A.~Balatsky,
\newblock \emph{Adv. Phys.} \textbf{2014}, \emph{63}, 1 1.

\bibitem{Ando_2013}
Y.~Ando,
\newblock \emph{J. Phys. Soc. Jpn.} \textbf{2013}, \emph{82}, 10 102001.

\bibitem{fu_2009}
L.~Fu,
\newblock \emph{Phys. Rev. Lett.} \textbf{2009}, \emph{103}, 26 266801.

\bibitem{Bardarson_2013}
J.~H. Bardarson, J.~E. Moore,
\newblock \emph{Rep. Progr. Phys.} \textbf{2013}, \emph{76}, 5 056501.

\bibitem{Bercioux_2015}
D.~Bercioux, P.~Lucignano,
\newblock \emph{Rep. Progr. Phys.} \textbf{2015}, \emph{78}, 10 106001.

\bibitem{Vergniory_2019}
M.~G. Vergniory, L.~Elcoro, C.~Felser, N.~Regnault, B.~A. Bernevig, Z.~Wang,
\newblock \emph{Nature} \textbf{2019}, \emph{566}, 7745 480.

\bibitem{Yang_2013}
W.-M. Yang, C.-J. Lin, J.~Liao, Y.-Q. Li,
\newblock \emph{Chin. Phys. B} \textbf{2013}, \emph{22}, 9 097202.

\bibitem{Culcer_2020}
D.~Culcer, A.~C. Keser, Y.~Li, G.~Tkachov,
\newblock \emph{2D Mater.} \textbf{2020}, \emph{7}, 2 022007.

\bibitem{Zhang_2011}
J.~Zhang, C.-Z. Chang, Z.~Zhang, J.~Wen, X.~Feng, K.~Li, M.~Liu, K.~He,
  L.~Wang, X.~Chen, Q.-K. Xue, X.~Ma, Y.~Wang,
\newblock \emph{Nat. Commun.} \textbf{2011}, \emph{2}, 1 574.

\bibitem{Kong_2011}
D.~Kong, Y.~Chen, J.~J. Cha, Q.~Zhang, J.~G. Analytis, K.~Lai, Z.~Liu, S.~S.
  Hong, K.~J. Koski, S.-K. Mo, Z.~Hussain, I.~R. Fisher, Z.-X. Shen, Y.~Cui,
\newblock \emph{Nat. Nanotechnol.} \textbf{2011}, \emph{6}, 11 705.

\bibitem{Burkov_2010}
A.~A. Burkov, D.~G. Hawthorn,
\newblock \emph{Phys. Rev. Lett.} \textbf{2010}, \emph{105}, 6 066802.

\bibitem{Mellnik_2014}
A.~R. Mellnik, J.~S. Lee, A.~Richardella, J.~L. Grab, P.~J. Mintun, M.~H.
  Fischer, A.~Vaezi, A.~Manchon, E.-A. Kim, N.~Samarth, D.~C. Ralph,
\newblock \emph{Nature} \textbf{2014}, \emph{511}, 7510 449.

\bibitem{Zhou2021}
X.~Zhou, W.-S. Lee, M.~Imada, N.~Trivedi, P.~Phillips, H.-Y. Kee,
  P.~T\"{o}rm\"{a}, M.~Eremets,
\newblock \emph{Nat. Rev. Phys.} \textbf{2021}, \emph{3}, 7 462.

\bibitem{Proust2019}
C.~Proust, L.~Taillefer,
\newblock \emph{Annu. Rev. Condens. Matter Phys.} \textbf{2019}, \emph{10}, 1
  409.

\bibitem{Keimer2015a}
B.~Keimer, S.~A. Kivelson, M.~R. Norman, S.~Uchida, J.~Zaanen,
\newblock \emph{Nature} \textbf{2015}, \emph{518}, 7538 179.

\bibitem{Hott2016}
R.~Hott, R.~Kleiner, T.~Wolf, G.~Zwicknagl,
\newblock \emph{Review on Superconducting Materials}, 1--55,
\newblock Wiley-VCH Verlag GmbH \& Co. KGaA, Weinheim, Germany, \textbf{2016}.

\bibitem{Kashiwaya2000}
S.~Kashiwaya, Y.~Tanaka,
\newblock \emph{Rep. Prog. Phys.} \textbf{2000}, \emph{63}, 10 1641.

\bibitem{Rogalla2011}
H.~Rogalla, P.~H. Kes,
\newblock \emph{100 years of superconductivity},
\newblock CRC Press, \textbf{2011}.

\bibitem{MacManus-Driscoll2021}
J.~L. MacManus-Driscoll, S.~C. Wimbush,
\newblock \emph{Nat. Rev. Mater.} \textbf{2021}, \emph{6}, 7 587.

\bibitem{Gupta_2018}
R.~Gupta, M.~Anerella, J.~Cozzolino, P.~Joshi, W.~Sampson, P.~Wanderer,
  J.~Kolonko, D.~Larson, R.~Scanlan, R.~Weggel, E.~Willen,
\newblock \emph{IEEE Trans. Appl. Supercond.} \textbf{2018}, \emph{28}, 3 1.

\bibitem{Lee1996}
A.~T. Lee, P.~L. Richards, S.~W. Nam, B.~Cabrera, K.~D. Irwin,
\newblock \emph{Appl. Phys. Lett.} \textbf{1996}, \emph{69}, 12 1801.

\bibitem{Kokkoniemi2020}
R.~Kokkoniemi, J.~P. Girard, D.~Hazra, A.~Laitinen, J.~Govenius, R.~E. Lake,
  I.~Sallinen, V.~Vesterinen, M.~Partanen, J.~Y. Tan, K.~W. Chan, K.~Y. Tan,
  P.~Hakonen, M.~M{\"{o}}tt{\"{o}}nen,
\newblock \emph{Nature} \textbf{2020}, \emph{586}, 7827 47.

\bibitem{Ullom2015a}
J.~N. Ullom, D.~A. Bennett,
\newblock \emph{Supercond. Sci. Technol.} \textbf{2015}, \emph{28}, 8 084003.

\bibitem{Crete2020}
D.~Cr{\'{e}}t{\'{e}}, Y.~Lematre, B.~Marcilhac, E.~Recoba-Pawlowski,
  J.~Trastoy, C.~Ulysse,
\newblock \emph{J. Phys. Conf. Ser} \textbf{2020}, \emph{1559}, 1 012012.

\bibitem{Couedo2019}
F.~Cou{\"e}do, E.~Recoba~Pawlowski, J.~Kermorvant, J.~Trastoy,
  D.~Cr{\'e}t{\'e}, Y.~Lema{\^\i}tre, B.~Marcilhac, C.~Ulysse,
  C.~Feuillet-Palma, N.~Bergeal, J.~Lesueur,
\newblock \emph{Appl. Phys. Lett.} \textbf{2019}, \emph{114}, 19 192602.

\bibitem{Shen1993}
Z.~Shen, D.~S. Dessau, B.~O. Wells, D.~M. King, W.~E. Spicer, A.~J. Arko,
  D.~Marshall, L.~W. Lombardo, A.~Kapitulnik, P.~Dickinson, S.~Doniach,
  J.~DiCarlo, T.~Loeser, C.~H. Park,
\newblock \emph{Phys. Rev. Lett.} \textbf{1993}, \emph{70}, 10 1553.

\bibitem{Aubin1997}
H.~Aubin, K.~Behnia, M.~Ribault, R.~Gagnon, L.~Taillefer,
\newblock \emph{Phys. Rev. Lett.} \textbf{1997}, \emph{78}, 13 2624.

\bibitem{Kirtley2006}
J.~R. Kirtley, C.~C. Tsuei, A.~Ariando, C.~J.~M. Verwijs, S.~Harkema,
  H.~Hilgenkamp,
\newblock \emph{Nat. Phys.} \textbf{2006}, \emph{2}, 3 190.

\bibitem{Kashiwaya_1995}
S.~Kashiwaya, Y.~Tanaka, M.~Koyanagi, H.~Takashima, K.~Kajimura,
\newblock \emph{Phys. Rev. B} \textbf{1995}, \emph{51}, 2 1350.

\bibitem{Wollman1993}
D.~A. Wollman, D.~J.~V. Harlingen, W.~C. Lee, D.~M. Ginsberg, A.~J. Leggett,
\newblock \emph{Phys. Rev. Lett.} \textbf{1993}, \emph{71}, 13 2134.

\bibitem{Tsuei1994}
C.~C. Tsuei, J.~R. Kirtley, C.~C. Chi, L.~S. Yu-Jahnes, A.~Gupta, T.~Shaw,
  J.~Z. Sun, M.~B. Ketchen,
\newblock \emph{Phys. Rev. Lett.} \textbf{1994}, \emph{73}, 4 593.

\bibitem{Grajcar1999}
M.~Grajcar, R.~Hlubina, R.~P. IJsselsteijn, H.~E. Hoenig, V.~Schultze, H.~G.
  Meyer,
\newblock \emph{Phys. Rev. B} \textbf{1999}, \emph{60}, 5 3096.

\bibitem{Bauch2005}
T.~Bauch, F.~Lombardi, F.~Tafuri, A.~Barone, G.~Rotoli, P.~Delsing, T.~Claeson,
\newblock \emph{Phys. Rev. Lett.} \textbf{2005}, \emph{94}, 8 087003.

\bibitem{Ariando2006}
{Ariando}, H.~J.~H. Smilde, C.~J.~M. Verwijs, G.~Rijnders, D.~H.~A. Blank,
  H.~Rogalla, J.~R. Kirtley, C.~C. Tsuei, H.~Hilgenkamp,
\newblock In \emph{Electron Correlation in New Materials and Nanosystems},
  149--174. Springer Netherlands, Dordrecht, \textbf{2007}.

\bibitem{Bauch2006}
T.~Bauch, T.~Lindstr{\"o}m, F.~Tafuri, G.~Rotoli, P.~Delsing, T.~Claeson,
  F.~Lombardi,
\newblock \emph{Science} \textbf{2006}, \emph{311}, 5757 57.

\bibitem{Cedergren2010}
K.~Cedergren, J.~R. Kirtley, T.~Bauch, G.~Rotoli, A.~Troeman, H.~Hilgenkamp,
  F.~Tafuri, F.~Lombardi,
\newblock \emph{Phys. Rev. Lett.} \textbf{2010}, \emph{104}, 17 177003.

\bibitem{Hilgenkamp2003}
H.~Hilgenkamp, {Ariando}, H.-J.~H. Smilde, D.~H.~A. Blank, G.~Rijnders,
  H.~Rogalla, J.~R. Kirtley, C.~C. Tsuei,
\newblock \emph{Nature} \textbf{2003}, \emph{422}, 6927 50.

\bibitem{Lucignano_2012}
P.~Lucignano, A.~Mezzacapo, F.~Tafuri, A.~Tagliacozzo,
\newblock \emph{Phys. Rev. B} \textbf{2012}, \emph{86}, 14 144513.

\bibitem{Lucignano_2013}
P.~Lucignano, F.~Tafuri, A.~Tagliacozzo,
\newblock \emph{Phys. Rev. B} \textbf{2013}, \emph{88}, 18 184512.

\bibitem{Trani_2016}
F.~Trani, G.~Campagnano, A.~Tagliacozzo, P.~Lucignano,
\newblock \emph{Phys. Rev. B} \textbf{2016}, \emph{94}, 13 134518.

\bibitem{Faley2020}
M.~I. Faley, P.~Reith, C.~D. Satrya, V.~S. Stolyarov, B.~Folkers, A.~A.
  Golubov, H.~Hilgenkamp,
\newblock \emph{Supercond. Sci. Technol.} \textbf{2020}, \emph{33}, 4 044005.

\bibitem{dean2010boron}
C.~R. Dean, A.~F. Young, I.~Meric, C.~Lee, L.~Wang, S.~Sorgenfrei, K.~Watanabe,
  T.~Taniguchi, P.~Kim, K.~L. Shepard, et~al.,
\newblock \emph{Nat. Nanotechnol.} \textbf{2010}, \emph{5}, 10 722.

\bibitem{Crassous2011}
A.~Crassous, R.~Bernard, S.~Fusil, K.~Bouzehouane, D.~Le~Bourdais,
  S.~Enouz-Vedrenne, J.~Briatico, M.~Bibes, A.~Barth{\'e}l{\'e}my, J.~E.
  Villegas,
\newblock \emph{Phys. Rev. Lett.} \textbf{2011}, \emph{107}, 24 247002.

\bibitem{Tolpygo1996}
S.~K. Tolpygo, J.~Lin, M.~Gurvitch, S.~Y. Hou, J.~M. Phillips,
\newblock \emph{Phys. Rev. B} \textbf{1996}, \emph{53}, 18 12462.

\bibitem{Trastoy2014}
J.~Trastoy, M.~Malnou, C.~Ulysse, R.~Bernard, N.~Bergeal, G.~Faini, J.~Lesueur,
  J.~Briatico, J.~E. Villegas,
\newblock \emph{Nat. Nanotechnol.} \textbf{2014}, \emph{9}, 9 710.

\bibitem{Yu2019}
Y.~Yu, L.~Ma, P.~Cai, R.~Zhong, C.~Ye, J.~Shen, G.~D. Gu, X.~H. Chen, Y.~Zhang,
\newblock \emph{Nature} \textbf{2019}, \emph{575}, 7781 156.

\bibitem{Sandilands2010}
L.~J. Sandilands, J.~X. Shen, G.~M. Chugunov, S.~Y. Zhao, S.~Ono, Y.~Ando,
  K.~S. Burch,
\newblock \emph{Phys. Rev. B} \textbf{2010}, \emph{82}, 6 1.

\bibitem{Zagoskin_2014}
A.~Zagoskin,
\newblock \emph{Quantum Theory of Many-Body Systems},
\newblock Springer International Publishing, \textbf{2014}.

\bibitem{Beenakker_2006}
C.~W.~J. Beenakker,
\newblock \emph{Phys. Rev. Lett.} \textbf{2006}, \emph{97}, 6 067007.

\bibitem{Efetov_2015}
D.~K. Efetov, L.~Wang, C.~Handschin, K.~B. Efetov, J.~Shuang, R.~Cava,
  T.~Taniguchi, K.~Watanabe, J.~Hone, C.~R. Dean, P.~Kim,
\newblock \emph{Nat. Phys.} \textbf{2015}, \emph{12}, 4 328.

\bibitem{Linder_2007}
J.~Linder, A.~Sudb{\o},
\newblock \emph{Phys. Rev. Lett.} \textbf{2007}, \emph{99}, 14 147001.

\bibitem{Linder_2008}
J.~Linder, A.~Sudb{\o},
\newblock \emph{Phys. Rev. B} \textbf{2008}, \emph{77}, 6 064507.

\bibitem{bruder1990andreev}
C.~Bruder,
\newblock \emph{Phys. Rev. B} \textbf{1990}, \emph{41}, 7 4017.

\bibitem{tanaka1995theory}
Y.~Tanaka, S.~Kashiwaya,
\newblock \emph{Phys. Rev. Lett.} \textbf{1995}, \emph{74}, 17 3451.

\bibitem{Blonder_1982}
G.~E. Blonder, M.~Tinkham, T.~M. Klapwijk,
\newblock \emph{Phys. Rev. B} \textbf{1982}, \emph{25}, 7 4515.

\bibitem{Chaudhuri_1995}
S.~Chaudhuri, P.~F. Bagwell,
\newblock \emph{Phys. Rev. B} \textbf{1995}, \emph{51}, 23 16936.

\bibitem{Mortensen_1999}
N.~A. Mortensen, K.~Flensberg, A.-P. Jauho,
\newblock \emph{Phys. Rev.B} \textbf{1999}, \emph{59}, 15 10176.

\bibitem{Hu_1994}
C.-R. Hu,
\newblock \emph{Phys. Rev. Lett.} \textbf{1994}, \emph{72}, 10 1526.

\bibitem{Kashiwaya_1996}
S.~Kashiwaya, Y.~Tanaka, M.~Koyanagi, K.~Kajimura,
\newblock \emph{Phys. Rev. B} \textbf{1996}, \emph{53}, 5 2667.

\bibitem{Bercioux_2018}
D.~Bercioux, P.~Lucignano,
\newblock \emph{Eur. Phys. J. ST} \textbf{2018}, \emph{227}, 12 1361.

\bibitem{Haake_2010}
F.~Haake,
\newblock \emph{Quantum Signatures of Chaos},
\newblock Springer Berlin Heidelberg, \textbf{2010}, in Chapter 2 it is
  explained in detail how to build up a TRS operator.

\bibitem{footnoteLWA}
The pairing potential in the LWA BdG Hamiltonian is obtained by performing the
  LWA on a complete tight-binding model in the particle-hole space.

\bibitem{Tkachov_2013_a}
G.~Tkachov, E.~M. Hankiewicz,
\newblock \emph{Phys. Status Solidi} \textbf{2013}, \emph{250}, 2 215.

\bibitem{Tkachov_2013_b}
G.~Tkachov, E.~M. Hankiewicz,
\newblock \emph{Phys. Rev. B} \textbf{2013}, \emph{88}, 7 075401.

\bibitem{Majidi_2016}
L.~Majidi, R.~Asgari,
\newblock \emph{Phys. Rev. B} \textbf{2016}, \emph{93}, 19 195404.

\bibitem{Gorkov_2001}
L.~P. Gorkov, E.~I. Rashba,
\newblock \emph{Phys. Rev. Lett.} \textbf{2001}, \emph{87}, 3 037004.

\bibitem{Burset_2015}
P.~Burset, B.~Lu, G.~Tkachov, Y.~Tanaka, E.~M. Hankiewicz, B.~Trauzettel,
\newblock \emph{Phys. Rev. B} \textbf{2015}, \emph{92}, 20 205424.

\bibitem{Breunig_2018}
D.~Breunig, P.~Burset, B.~Trauzettel,
\newblock \emph{Phys. Rev. Lett.} \textbf{2018}, \emph{120}, 3 037701.

\bibitem{Qi_2009}
X.-L. Qi, T.~L. Hughes, S.~Raghu, S.-C. Zhang,
\newblock \emph{Phys. Rev. Lett.} \textbf{2009}, \emph{102}, 18 187001.

\bibitem{Fu2008}
L.~Fu, C.~L. Kane,
\newblock \emph{Phys. Rev. Lett.} \textbf{2008}, \emph{100} 096407.

\bibitem{Tanaka_2009}
Y.~Tanaka, T.~Yokoyama, N.~Nagaosa,
\newblock \emph{Phys. Rev. Lett.} \textbf{2009}, \emph{103}, 10 107002.

\bibitem{Law_2009}
K.~T. Law, P.~A. Lee, T.~K. Ng,
\newblock \emph{Phys. Rev. Lett.} \textbf{2009}, \emph{103}, 23 237001.

\bibitem{Katsnelson_2006}
M.~I. Katsnelson, K.~S. Novoselov, A.~K. Geim,
\newblock \emph{Nat. Phys.} \textbf{2006}, \emph{2}, 9 620.

\bibitem{Young_2009}
A.~F. Young, P.~Kim,
\newblock \emph{Nat. Phys.} \textbf{2009}, \emph{5}, 3 222.

\bibitem{Tudorovskiy_2012}
T.~Tudorovskiy, K.~J.~A. Reijnders, M.~I. Katsnelson,
\newblock \emph{Phys. Scr.} \textbf{2012}, \emph{T146} 014010.

\bibitem{Bhattacharjee_2006}
S.~Bhattacharjee, K.~Sengupta,
\newblock \emph{Phys. Rev. Lett.} \textbf{2006}, \emph{97}, 21 217001.

\bibitem{HLi_2017}
H.~Li, Y.~Y. Zhao,
\newblock \emph{J. Phys. Condens. Matter} \textbf{2017}, \emph{29}, 46 465001.

\bibitem{Hu_1998}
C.-R. Hu,
\newblock \emph{Phys. Rev. B} \textbf{1998}, \emph{57}, 2 1266.

\bibitem{Black_Schaffer_2007}
A.~M. Black-Schaffer, S.~Doniach,
\newblock \emph{Phys. Rev. B} \textbf{2007}, \emph{75}, 13 134512.

\bibitem{Black-Schaffer_2014}
A.~M. Black-Schaffer, C.~Honerkamp,
\newblock \emph{J. Phys. Condens. Matter} \textbf{2014}, \emph{26}, 42 423201.

\bibitem{Linder_2009}
J.~Linder, A.~M. Black-Schaffer, T.~Yokoyama, S.~Doniach, A.~Sudb{\o},
\newblock \emph{Phys. Rev. B} \textbf{2009}, \emph{80}, 9 094522.

\bibitem{Linder_2010}
J.~Linder, Y.~Tanaka, T.~Yokoyama, A.~Sudb{\o}, N.~Nagaosa,
\newblock \emph{Phys. Rev. Lett.} \textbf{2010}, \emph{104}, 6 067001.

\bibitem{Sato_2009}
M.~Sato, S.~Fujimoto,
\newblock \emph{Phys. Rev. B} \textbf{2009}, \emph{79}, 9 094504.

\bibitem{Zhang_2013}
F.~Zhang, C.~L. Kane, E.~J. Mele,
\newblock \emph{Phys. Rev. Lett.} \textbf{2013}, \emph{111}, 5 056402.

\bibitem{Salehi_2010}
M.~Salehi, M.~Alidoust, G.~Rashedi,
\newblock \emph{J. Appl. Phys.} \textbf{2010}, \emph{108}, 8 083917.

\bibitem{Dean}
C.~R. Dean, A.~F. Young, I.~Meric, C.~Lee, L.~Wang, S.~Sorgenfrei, K.~Watanabe,
  T.~Taniguchi, P.~Kim, K.~L. Shepard, J.~Hone,
\newblock \emph{Nat. Nanotechnol.} \textbf{2010}, \emph{5}, 10 722.

\bibitem{Lee2017}
G.-H. Lee, K.-F. Huang, D.~K. Efetov, D.~S. Wei, S.~Hart, T.~Taniguchi,
  K.~Watanabe, A.~Yacoby, P.~Kim,
\newblock \emph{Nat. Phys.} \textbf{2017}, \emph{13}, 7 693.

\bibitem{Mong2014}
R.~S. Mong, D.~J. Clarke, J.~Alicea, N.~H. Lindner, P.~Fendley, C.~Nayak,
  Y.~Oreg, A.~Stern, E.~Berg, K.~Shtengel, M.~P. Fisher,
\newblock \emph{Phys. Rev. X} \textbf{2014}, \emph{4}, 1 011036.

\bibitem{Bonaccorso2012}
F.~Bonaccorso, A.~Lombardo, T.~Hasan, Z.~Sun, L.~Colombo, A.~C. Ferrari,
\newblock \emph{Mater. Today} \textbf{2012}, \emph{15}, 12 564.

\bibitem{Jiang2008}
Y.~Jiang, D.-X. Yao, E.~W. Carlson, H.-D. Chen, J.~Hu,
\newblock \emph{Phys. Rev. B} \textbf{2008}, \emph{77}, 23 235420.

\bibitem{Sun2014a}
Q.~J. Sun, H.~S. Wang, H.~M. Wang, L.~W. Deng, Z.~W. Hu, B.~Gao, Q.~Li, X.~M.
  Xie,
\newblock \emph{Appl. Phys. Lett.} \textbf{2014}, \emph{104}, 10 102602.

\bibitem{Chang2004}
H.~S. Chang, M.~H. Bae, H.~J. Lee,
\newblock \emph{Physica C Supercond.} \textbf{2004}, \emph{408}, 1-4 618.

\bibitem{Tsuneto_1962}
T.~Tsuneto,
\newblock \emph{Prog. Theor. Exp. Phys.} \textbf{1962}, \emph{28}, 5 857.

\bibitem{Balatsky_2006}
A.~V. Balatsky, I.~Vekhter, J.-X. Zhu,
\newblock \emph{Rev. Mod. Phys.} \textbf{2006}, \emph{78}, 2 373.

\bibitem{Huard2007}
B.~Huard, J.~A. Sulpizio, N.~Stander, K.~Todd, B.~Yang, D.~Goldhaber-Gordon,
\newblock \emph{Phys. Rev. Lett.} \textbf{2007}, \emph{98}, 23 236803.

\bibitem{de_Gennes_1963}
P.~de~Gennes, D.~Saint-James,
\newblock \emph{Phys. Lett.} \textbf{1963}, \emph{4}, 2 151.

\bibitem{Rowell_1966}
J.~M. Rowell, W.~L. McMillan,
\newblock \emph{Phys. Rev. Lett.} \textbf{1966}, \emph{16}, 11 453.

\bibitem{Liang_2001}
W.~Liang, M.~Bockrath, D.~Bozovic, J.~H. Hafner, M.~Tinkham, H.~Park,
\newblock \emph{Nature} \textbf{2001}, \emph{411}, 6838 665.

\bibitem{Miao_2007}
F.~Miao, S.~Wijeratne, Y.~Zhang, U.~C. Coskun, W.~Bao, C.~N. Lau,
\newblock \emph{Science} \textbf{2007}, \emph{317}, 5844 1530.

\bibitem{Campos_2012}
L.~Campos, A.~Young, K.~Surakitbovorn, K.~Watanabe, T.~Taniguchi,
  P.~Jarillo-Herrero,
\newblock \emph{Nat. Comm.} \textbf{2012}, \emph{3}, 1 1239.

\bibitem{Allen_2017}
M.~T. Allen, O.~Shtanko, I.~C. Fulga, J.~I.-J. Wang, D.~Nurgaliev, K.~Watanabe,
  T.~Taniguchi, A.~R. Akhmerov, P.~Jarillo-Herrero, L.~S. Levitov, A.~Yacoby,
\newblock \emph{Nano Lett.} \textbf{2017}, \emph{17}, 12 7380.

\bibitem{Wei_1998}
J.~Y.~T. Wei, N.-C. Yeh, D.~F. Garrigus, M.~Strasik,
\newblock \emph{Phys. Rev. Lett.} \textbf{1998}, \emph{81}, 12 2542.

\bibitem{Xu2014}
S.~Y. Xu, C.~Liu, A.~Richardella, I.~Belopolski, N.~Alidoust, M.~Neupane,
  G.~Bian, N.~Samarth, M.~Z. Hasan,
\newblock \emph{Phys. Rev. B} \textbf{2014}, \emph{90}, 8 085128.

\bibitem{Zareapour2012}
P.~Zareapour, A.~Hayat, S.~Y.~F. Zhao, M.~Kreshchuk, A.~Jain, D.~C. Kwok,
  N.~Lee, S.-W. Cheong, Z.~Xu, A.~Yang, G.~D. Gu, S.~Jia, R.~J. Cava, K.~S.
  Burch,
\newblock \emph{Nat. Commun.} \textbf{2012}, \emph{3}, 1 1056.

\bibitem{Qi2011}
X.~L. Qi, S.~C. Zhang,
\newblock \emph{Rev. Mod. Phys.} \textbf{2011}, \emph{83}, 4 1057.

\bibitem{Beenakker2015}
C.~W. Beenakker,
\newblock \emph{Rev. Mod. Phys.} \textbf{2015}, \emph{87}, 3 1037.

\bibitem{Asano2009}
Y.~Asano, I.~Suemune, H.~Takayanagi, E.~Hanamura,
\newblock \emph{Phys. Rev. Lett.} \textbf{2009}, \emph{103}, 18 187001.

\bibitem{Hayat2012}
A.~Hayat, P.~Zareapour, S.~Y.~F. Zhao, A.~Jain, I.~G. Savelyev, M.~Blumin,
  Z.~Xu, A.~Yang, G.~D. Gu, H.~E. Ruda, S.~Jia, R.~J. Cava, A.~M. Steinberg,
  K.~S. Burch,
\newblock \emph{Phys. Rev. X} \textbf{2012}, \emph{2}, 4 041019.

\bibitem{Hayat2014}
A.~Hayat, H.-Y. Kee, K.~S. Burch, A.~M. Steinberg,
\newblock \emph{Phys. Rev. B} \textbf{2014}, \emph{89}, 9 094508.

\bibitem{Bocquillon2017}
E.~Bocquillon, R.~S. Deacon, J.~Wiedenmann, P.~Leubner, T.~M. Klapwijk,
  C.~Br{\"{u}}ne, K.~Ishibashi, H.~Buhmann, L.~W. Molenkamp,
\newblock \emph{Nat. Nanotechnol.} \textbf{2017}, \emph{12}, 2 137.

\bibitem{Bocquillon2018}
E.~Bocquillon, J.~Wiedenmann, R.~S. Deacon, T.~M. Klapwijk, H.~Buhmann, L.~W.
  Molenkamp,
\newblock In \emph{Topological Matter}, 115--148. Springer International
  Publishing, \textbf{2018}.

\bibitem{Sacepe2011}
B.~Sac{\'e}p{\'e}, J.~B. Oostinga, J.~Li, A.~Ubaldini, N.~J.~G. Couto,
  E.~Giannini, A.~F. Morpurgo,
\newblock \emph{Nat. Commun.} \textbf{2011}, \emph{2}, 1 575.

\bibitem{Kurter2014}
C.~Kurter, Y.~S. Hor, D.~J.~V. Harlingen,
\newblock \emph{Phys. Rev. X} \textbf{2014}, \emph{4}, 4 041022.

\bibitem{kunakova2020topological}
G.~Kunakova, A.~P. Surendran, D.~Montemurro, M.~Salvato, D.~Golubev,
  J.~Andzane, D.~Erts, T.~Bauch, F.~Lombardi,
\newblock \emph{J. Appl. Phys.} \textbf{2020}, \emph{128}, 19 194304.

\bibitem{charpentier2017induced}
S.~Charpentier, L.~Galletti, G.~Kunakova, R.~Arpaia, Y.~Song, R.~Baghdadi,
  S.~M. Wang, A.~Kalaboukhov, E.~Olsson, F.~Tafuri, D.~Golubev, J.~Linder,
  T.~Bauch, F.~Lombardi,
\newblock \emph{Nat. Commun.} \textbf{2017}, \emph{8}, 1.

\bibitem{galletti2014influence}
L.~Galletti, S.~Charpentier, M.~Iavarone, P.~Lucignano, D.~Massarotti,
  R.~Arpaia, Y.~Suzuki, K.~Kadowaki, T.~Bauch, A.~Tagliacozzo, et~al.,
\newblock \emph{Phys. Rev. B} \textbf{2014}, \emph{89}, 13 134512.

\bibitem{rosen2021fractional}
I.~T. Rosen, C.~J. Trimble, M.~P. Andersen, E.~Mikheev, Y.~Li, Y.~Liu, L.~Tai,
  P.~Zhang, K.~L. Wang, Y.~Cui, M.~A. Kastner, J.~R. Williams,
  D.~Goldhaber-Gordon,
\newblock \emph{arXiv preprint arXiv:2110.01039} \textbf{2021}.

\bibitem{Wang2018}
A.-Q. Wang, C.-Z. Li, C.~Li, Z.-M. Liao, A.~Brinkman, D.-P. Yu,
\newblock \emph{Phys. Rev. Lett.} \textbf{2018}, \emph{121}, 23 237701.

\bibitem{Yu2020}
W.~Yu, R.~Haenel, M.~A. Rodriguez, S.~R. Lee, F.~Zhang, M.~Franz, D.~I.
  Pikulin, W.~Pan,
\newblock \emph{Phys. Rev. Research} \textbf{2020}, \emph{2}, 3 032002.

\bibitem{li20184pi}
C.~Li, J.~C. de~Boer, B.~de~Ronde, S.~V. Ramankutty, E.~van Heumen, Y.~Huang,
  A.~de~Visser, A.~A. Golubov, M.~S. Golden, A.~Brinkman,
\newblock \emph{Nat. Mat.} \textbf{2018}, \emph{17}, 10 875.

\bibitem{li2019zeeman}
C.~Li, B.~de~Ronde, J.~de~Boer, J.~Ridderbos, F.~Zwanenburg, Y.~Huang,
  A.~Golubov, A.~Brinkman,
\newblock \emph{Phys. Rev. Lett.} \textbf{2019}, \emph{123}, 2 026802.

\bibitem{Zareapour2014}
P.~Zareapour, A.~Hayat, S.~Y.~F. Zhao, M.~Kreshchuk, Y.~K. Lee, A.~A.
  Reijnders, A.~Jain, Z.~Xu, T.~S. Liu, G.~D. Gu, S.~Jia, R.~J. Cava, K.~S.
  Burch,
\newblock \emph{Phys. Rev. B} \textbf{2014}, \emph{90}, 24 241106.

\bibitem{Yilmaz2014}
T.~Yilmaz, I.~Pletikosi{\'{c}}, A.~P. Weber, J.~T. Sadowski, G.~D. Gu, A.~N.
  Caruso, B.~Sinkovic, T.~Valla,
\newblock \emph{Phys. Rev. Lett.} \textbf{2014}, \emph{113}, 6 067003.

\bibitem{Zareapour2016}
P.~Zareapour, J.~Xu, S.~Y.~F. Zhao, A.~Jain, Z.~Xu, T.~S. Liu, G.~D. Gu, K.~S.
  Burch,
\newblock \emph{Supercond. Sci. Technol.} \textbf{2016}, \emph{29}, 12 125006.

\bibitem{Fei2017}
Z.~Fei, T.~Palomaki, S.~Wu, W.~Zhao, X.~Cai, B.~Sun, P.~Nguyen, J.~Finney,
  X.~Xu, D.~H. Cobden,
\newblock \emph{Nat. Phys.} \textbf{2017}, \emph{13}, 7 677.

\bibitem{Ma2019}
Q.~Ma, S.~Y. Xu, H.~Shen, D.~MacNeill, V.~Fatemi, T.~R. Chang, A.~M. {Mier
  Valdivia}, S.~Wu, Z.~Du, C.~H. Hsu, S.~Fang, Q.~D. Gibson, K.~Watanabe,
  T.~Taniguchi, R.~J. Cava, E.~Kaxiras, H.~Z. Lu, H.~Lin, L.~Fu, N.~Gedik,
  P.~Jarillo-Herrero,
\newblock \emph{Nature} \textbf{2019}, \emph{565}, 7739 337.

\bibitem{Fatem}
V.~Fatemi, S.~Wu, Y.~Cao, L.~Bretheau, Q.~Gibson, K.~Watanabe, T.~Taniguchi,
  R.~J. Cava, P.~Jarillo-Herrero,
\newblock \emph{Science} \textbf{2018}, \emph{362}, 6417 926.

\bibitem{Sajadi2018}
E.~Sajadi, T.~Palomaki, Z.~Fei, W.~Zhao, P.~Bement, C.~Olsen, S.~Luescher,
  X.~Xu, J.~A. Folk, D.~H. Cobden,
\newblock \emph{Science} \textbf{2018}, \emph{362}, 6417 922.

\bibitem{DeVries2018a}
F.~K. {De Vries}, T.~Timmerman, V.~P. Ostroukh, J.~{Van Veen}, A.~J. Beukman,
  F.~Qu, M.~Wimmer, B.~M. Nguyen, A.~A. Kiselev, W.~Yi, M.~Sokolich, M.~J.
  Manfra, C.~M. Marcus, L.~P. Kouwenhoven,
\newblock \emph{Phys. Rev. Lett.} \textbf{2018}, \emph{120}, 4 047702.

\bibitem{Liu2016}
Y.~Liu, N.~O. Weiss, X.~Duan, H.-C. Cheng, Y.~Huang, X.~Duan,
\newblock \emph{Nat. Rev. Mater.} \textbf{2016}, \emph{1}, 9.

\bibitem{Novoselov2016}
K.~Novoselov, A.~Mishchenko, A.~Carvalho, A.~H. {Castro Neto},
\newblock \emph{Science} \textbf{2016}, \emph{353}, 6298 1.

\bibitem{Zhu2021}
Y.~Zhu, M.~Liao, Q.~Zhang, H.~Y. Xie, F.~Meng, Y.~Liu, Z.~Bai, S.~Ji, J.~Zhang,
  K.~Jiang, R.~Zhong, J.~Schneeloch, G.~Gu, L.~Gu, X.~Ma, D.~Zhang, Q.~K. Xue,
\newblock \emph{Phys. Rev. X} \textbf{2021}, \emph{11}, 3 31011.

\bibitem{Zhao2021}
S.~Y.~F. Zhao, N.~Poccia, X.~Cui, P.~A. Volkov, H.~Yoo, R.~Engelke, Y.~Ronen,
  R.~Zhong, G.~Gu, S.~Plugge, T.~Tummuru, M.~Franz, J.~H. Pixley, P.~Kim,
\newblock \emph{arxiv:2108.13455v1} \textbf{2021}.

\end{thebibliography}

\begin{figure}
  \includegraphics[width=2cm]{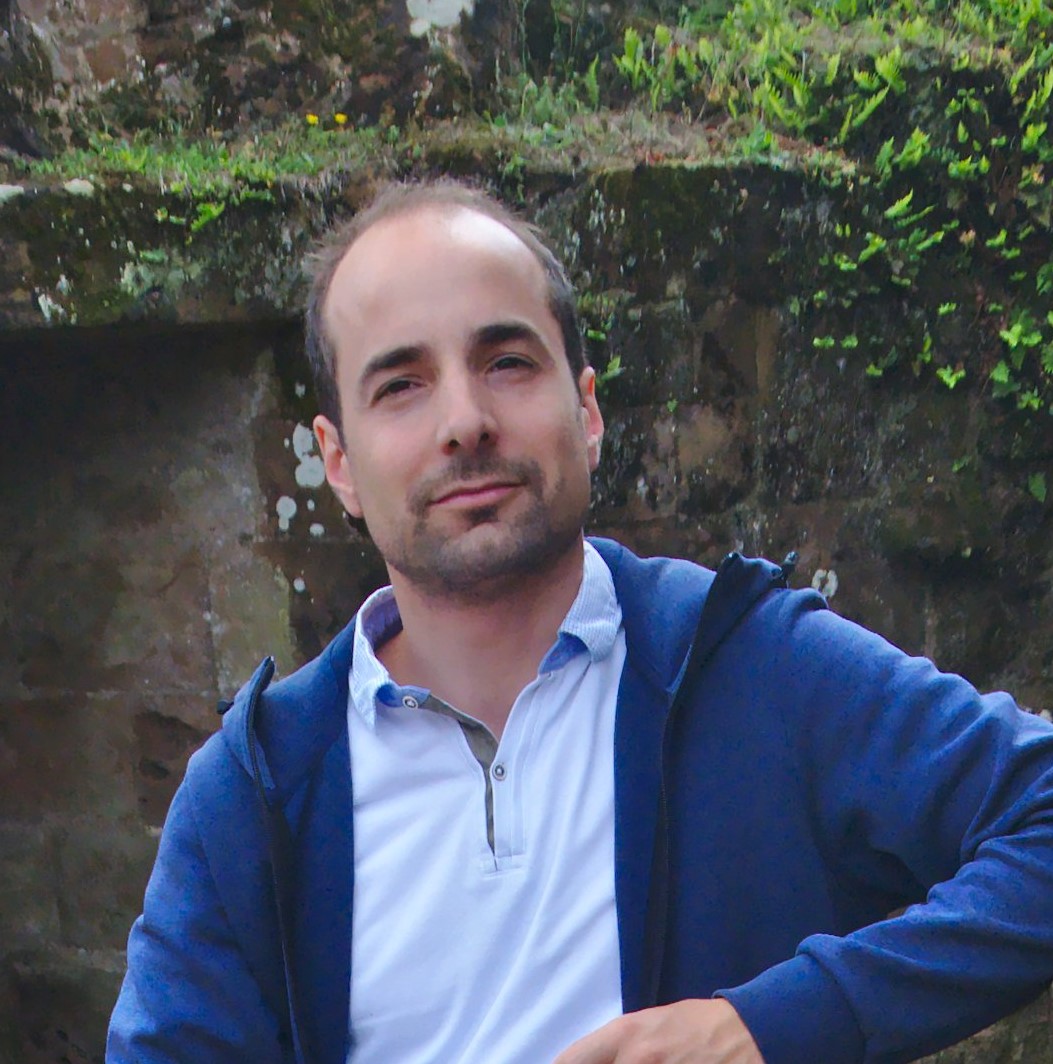}
  \caption*{David Perconte is a postdoctoral Researcher at the institut Néel lab in Grenoble. He received his PhD from the University Pierre et Marie Curie in 2018 working on graphene d-wave superconductor junction. He later did a postdoctoral stay in the Autonoma university of Madrid working in the group of Isabel Guillamon and Hermann Suderow.}
\end{figure}
\begin{figure}
  \includegraphics[width=2cm]{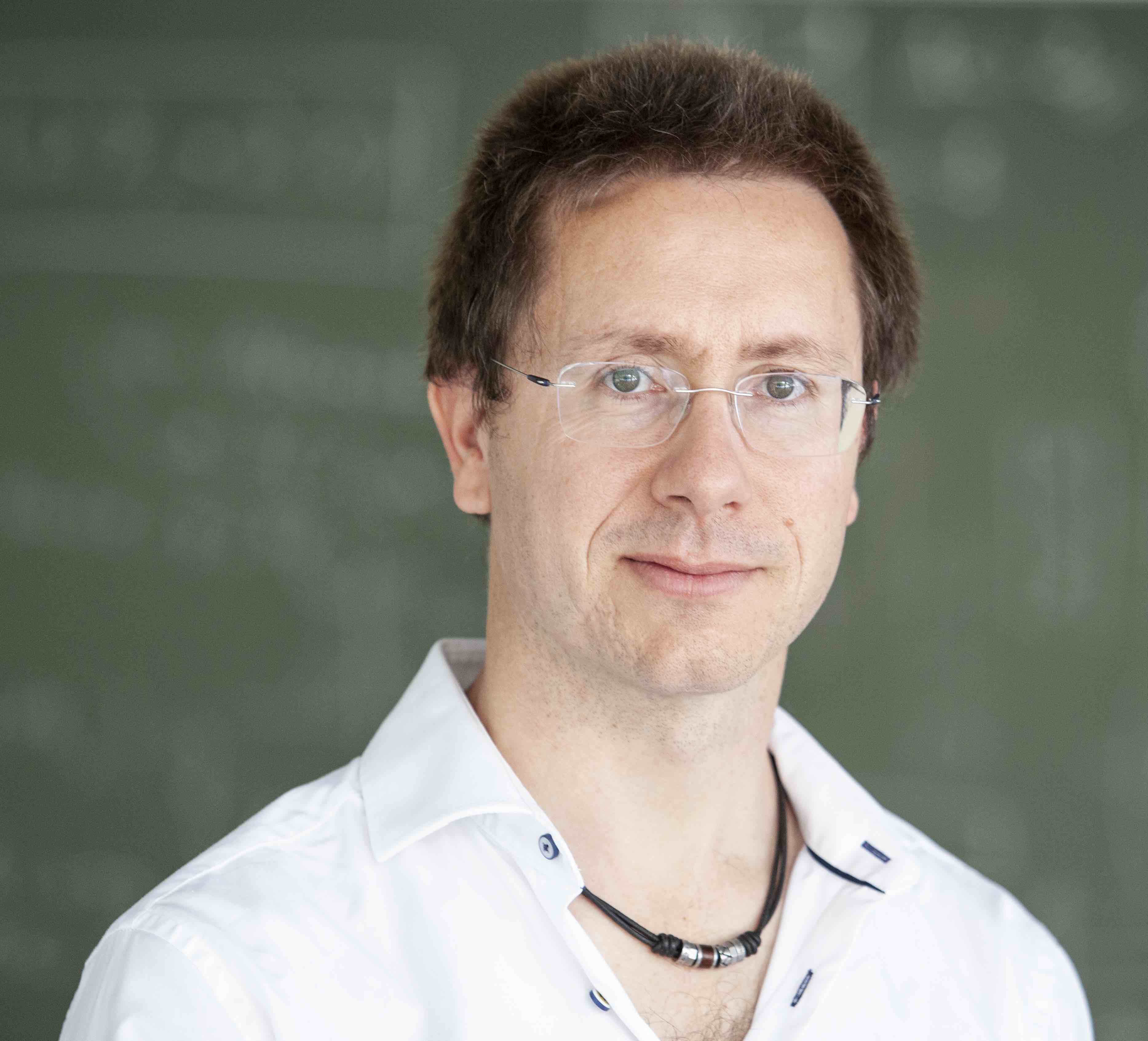}
  \caption*{Dario Bercioux is an Ikerbasque Research Associate at the Donostia International Physics Center (DIPC) in Donostia-San Sebasti\'an. He received is PhD from the University of Naples in 2005. Afterword he was a postdoctoral fellow in Regensburg, Freiburg and at the FU Berlin. He joined the DIPC in the 2014 as Ikerbasque Research Fellow.}
\end{figure}
\begin{figure}
  \includegraphics[width=2cm]{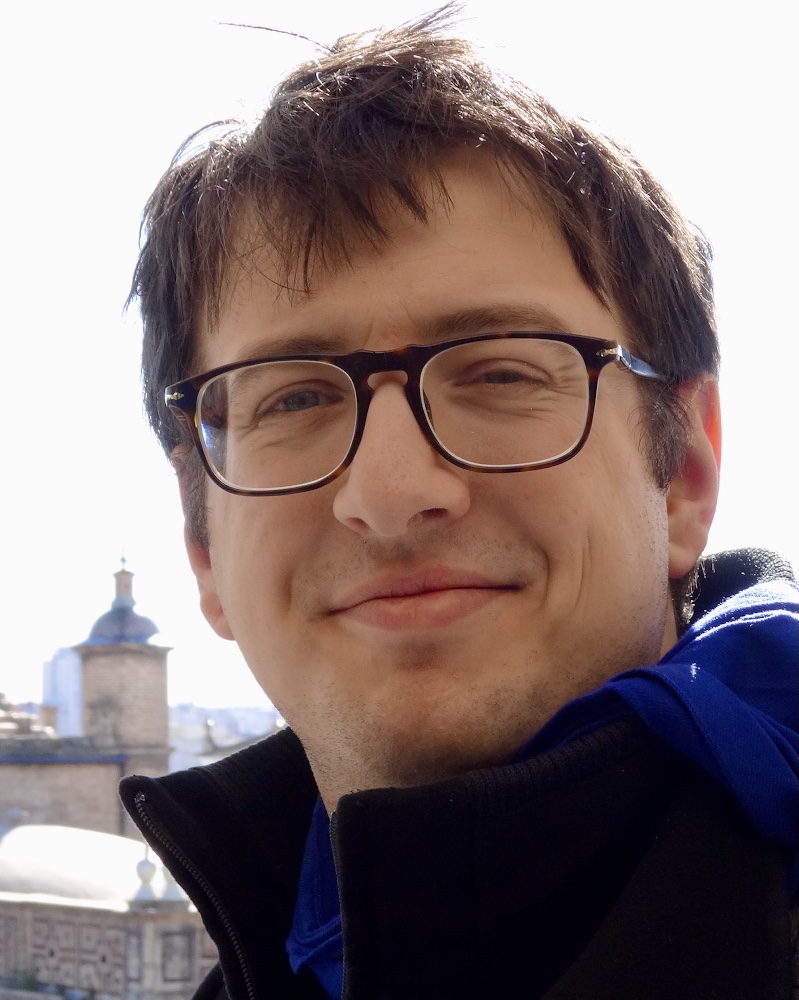}
  \caption*{Bruno Dlubak is a CNRS Researcher at the Unité Mixte de Physique CNRS-Thales lab, associated to the Paris-Saclay University. He received his PhD from the University of Paris-Sud in 2011 working on graphene spintronics. After a postdoctoral appointment at the University of Cambridge working in John Robertson's group, he became a CNRS Researcher in 2014.}
\end{figure}
\begin{figure}
  \includegraphics[width=2cm]{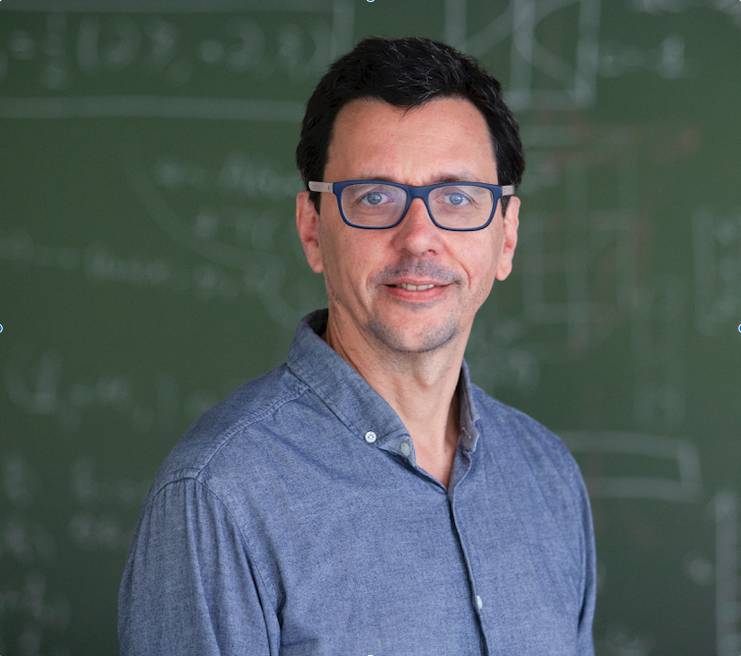}
  \caption*{F. Sebastian Bergeret is a Researcher at the Materials Physics Center (CSIC) and  head of the Mesoscopic Physics Group where he leads research on quantum transport, superconductivity and magnetism. He earned his PhD at the Bochum University  (Germany) in 2002.}
\end{figure}
\begin{figure}
  \includegraphics[width=2cm]{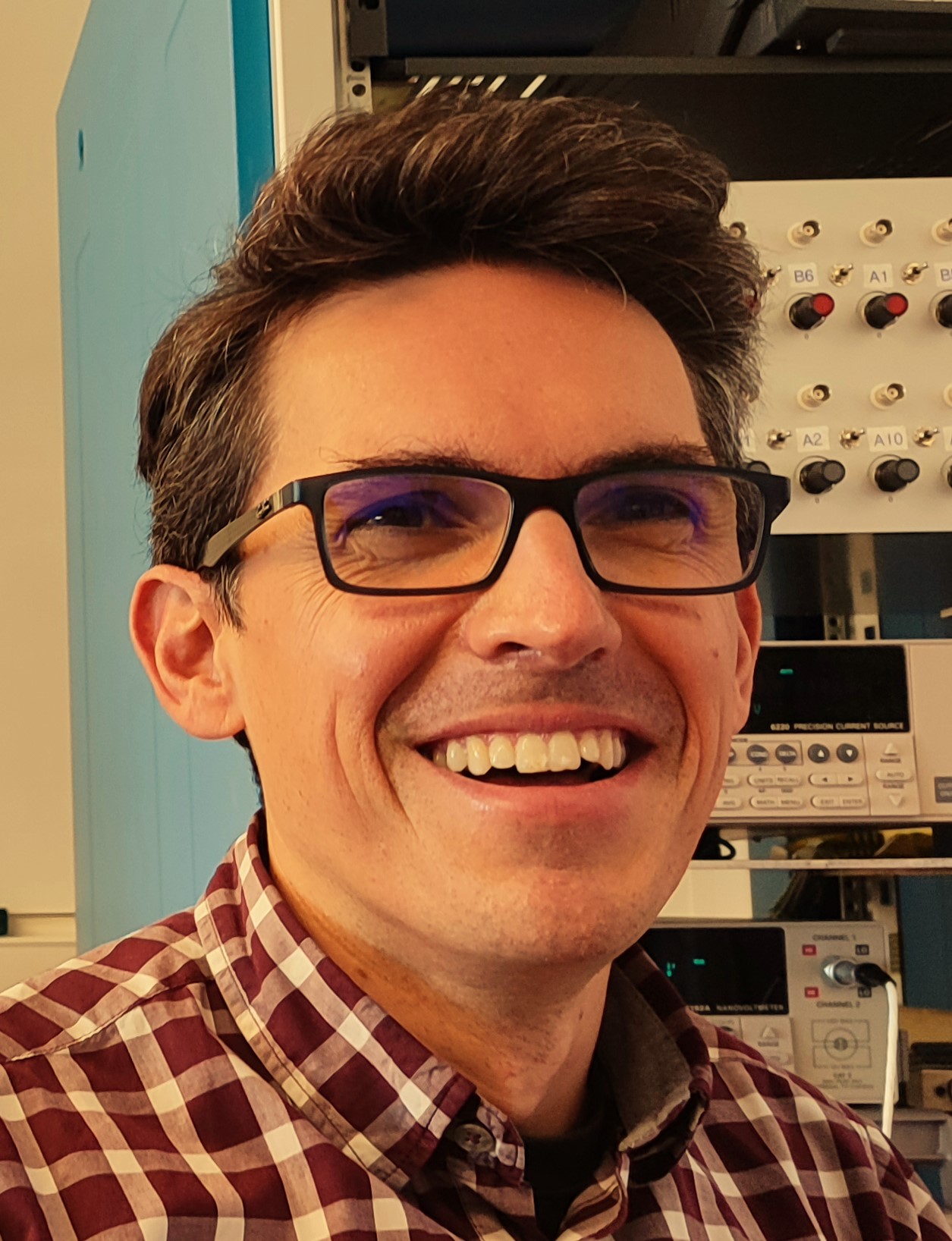}
  \caption*{Javier E. Villegas is a Research Director at the CNRS. He leads research on superconducting hybrids and oxides the CNRS/Thales joint lab (France). He earned his PhD at the University Complutense of Madrid (Spain) in 2004, where he also worked as adjunct professor until 2005. Following a postdoctoral research period in the group of I.K. Schuller at the University of California-San Diego (USA), he became CNRS permanent staff in 2007.}
\end{figure}





\end{document}